\documentclass[%
 reprint,
superscriptaddress,
 amsmath,amssymb,
 longbibliography,
 prb,
]{revtex4-2}

\usepackage[ruled]{algorithm2e}
\usepackage[noend]{algpseudocode}
\usepackage{wasysym}
\usepackage{graphicx}
\usepackage{dcolumn}
\usepackage{bm}
\DeclareMathAlphabet{\mathpzc}{OT1}{pzc}{m}{it}

\usepackage{color}

\begin{document}

\preprint{APS/123-QED}

\title{Generalized Lanczos method for systematic optimization of neural-network quantum states}

\author{Jia-Qi Wang}
\affiliation{Department of Physics, Renmin University of China, Beijing 100872, China}
\affiliation{Key Laboratory of Quantum State Construction and Manipulation (Ministry of Education), Renmin University of China, Beijing 100872, China}

\author{Rong-Qiang He}\email{rqhe@ruc.edu.cn}
\affiliation{Department of Physics, Renmin University of China, Beijing 100872, China}
\affiliation{Key Laboratory of Quantum State Construction and Manipulation (Ministry of Education), Renmin University of China, Beijing 100872, China}

\author{Zhong-Yi Lu}\email{zlu@ruc.edu.cn}
\affiliation{Department of Physics, Renmin University of China, Beijing 100872, China}
\affiliation{Key Laboratory of Quantum State Construction and Manipulation (Ministry of Education), Renmin University of China, Beijing 100872, China}
\affiliation{Hefei National Laboratory, Hefei 230088, China}

\date{\today}

\begin{abstract}
  Recently, artificial intelligence for science has made significant inroads into various fields of natural science research. In the field of quantum many-body computation, researchers have developed numerous ground state solvers based on neural-network quantum states (NQSs), achieving ground state energies with accuracy comparable to or surpassing traditional methods such as variational Monte Carlo methods, density matrix renormalization group, and quantum Monte Carlo methods. Here, we combine supervised learning, variational Monte Carlo (VMC), and the Lanczos method to develop a systematic approach to improving the NQSs of many-body systems, which we refer to as the NQS Lanczos method. The algorithm mainly consists of two parts: the supervised learning part and the VMC optimization part. Through supervised learning, the Lanczos states are represented by the NQSs. Through VMC, the NQSs are further optimized. We analyze the reasons for the underfitting problem and demonstrate how the NQS Lanczos method systematically improves the energy in the highly frustrated regime of the two-dimensional Heisenberg $J_1$-$J_2$ model. Compared to the existing method that combines the Lanczos method with the restricted Boltzmann machine, the primary advantage of the NQS Lanczos method is its linearly increasing computational cost.
\end{abstract}


\maketitle


\section{Introduction}
\label{sec:introduction}
The ground state of a quantum many-body system has long been one of the central topics in condensed matter physics. It is crucial for understanding fundamental physical properties. When the number of particles is large, the existing methods fail to find the exact ground state due to the ``exponential wall'' problem.

To obtain the ground state of a many-body system, many methods have been developed in the past, including the variational Monte Carlo (VMC) methods~\cite{Sandvik2007VBBasis,Sandvik2010_VBBasis+LoopUpdate,Sorella2013_VMC,Sandvik2007_VMC+TNS,Tagliacozzo2009_TreeTensorNetwork}, density matrix renormalization group~\cite{GongShouShu2014_DMRG,Stoudenmire2012_2DsystemDMRG,Sandvik2018_DMRG}, and the quantum Monte Carlo methods~\cite{Sandvik1997SSEQMC,Sandvik2010_VBBasis+LoopUpdate}, and have effectively advanced the field. In recent years, with the increasing use of artificial intelligence technologies in quantum many-body physics, researchers have started using neural networks to represent many-body wave functions, known as neural-network quantum states (NQSs). NQS-based methods are being rapidly developed~\cite{Juan2021PRX_NQSReview,Nomura2023RBMNQSReview,Hermann2023_NQSChemistryReview,Medvidovic2024NQSReview,Lange2024NQSReview}, whose performances are even surpassing those of traditional approaches.

NQSs were first introduced in 2017 by Carleo and Troyer~\cite{GiuseppeCarleo2017_RBM}, who applied restricted Boltzmann machines (RBMs)  with stochastic reconfiguration (SR) optimization~\cite{Sorella2007SR} to the field of quantum many-body computation. Since then, researchers have begun to study the problems~\cite{Claudio2020_NQS_SignProblem,westerhout2023SignStructure,Westerhout2020_difference_Sign_amp,Chen2022_SignStructure-NeuralNetwork} encountered in the optimization process and try to use new network architectures~\cite{Cai2018_ANN,Pfau2020Ferminet,HibatAllah2020_RNN_NQS,kochkov2021GNN,fu2022latticeCNN,roth2023_GCNN,Rende2024DeepViT,Lange2024RNN_NQS} and optimization algorithms~\cite{Chen2024MinSR,Drissi2024SecondOrderOptimization}. For example, Pfau \textit{et al.}~\cite{Pfau2020Ferminet} incorporated the antisymmetry of the exchange of fermions into the neural network and gave a ground state solver for small molecular systems. Westerhout \textit{et al.}~\cite{Westerhout2020_difference_Sign_amp} studied the generalization capabilities of NQSs in the Hilbert space and explored the optimization of both the amplitude and sign of the many-body wave functions for frustrated systems. Roth \textit{et al.}~\cite{roth2023_GCNN} incorporated group symmetries into convolutional neural networks (CNNs), enhancing the ability of CNN to handle two-dimensional lattice systems. Chen and Heyl~\cite{Chen2024MinSR} introduced the minimum-step stochastic reconfiguration (MinSR) algorithm, an improvement on the SR algorithm based on imaginary time evolution. The accuracy of the ground state energy obtained by the test (on the two-dimensional Heisenberg model with $J_2/J_1 = 0.5$) in Ref.~\cite{Chen2024MinSR} surpassed the one by traditional methods, highlighting the significant potential of NQSs in solving quantum many-body problems. However, the trade-off is that the network parameters quickly grow to millions. For example, in the latest simulations of the 10 × 10 square-lattice Heisenberg $J_1$-$J_2$ model, Rende \textit{et al.}~\cite{Rende2024DeepViT} used 267,720 parameters, while Chen and Heyl~\cite{Chen2024MinSR} used 1,071,488 parameters. The approach of increasing the number of network parameters is not sustainable. As the network size increases, the computational cost increases dramatically, while the returns diminish rapidly.

The Lanczos method~\cite{Lanczos1952Method} is an approach to obtain accurate ground state or low-lying excitations of small quantum systems. It was combined with VMC methods~\cite{sorella2001VMC_Lanczos} and tensor networks~\cite{huang2018tensornetwork_Lanczos} for treating larger systems. The Lanczos method starts from an arbitrary state (which is not orthogonal to the ground state) and progressively constructs new states that are orthogonal to the previous states in the Krylov space. An orthogonal basis set can be constructed by these Lanczos states. The Hamiltonian matrix in this basis becomes tridiagonal. By diagonalizing it, an approximate ground state, represented as a superposition of these basis states, is obtained. The corresponding eigenvalue gives the approximate ground state energy.

In 2022, Chen \textit{et al.}~\cite{chen2022lanczos} applied the Lanczos method to two-dimensional lattice systems, where wave functions are represented by RBMs, successfully improving the energies. However, this method needs to calculate the expectations $\langle H^{2i + 1} \rangle$, where $H$ is the Hamiltonian, causing the computational cost to increase exponentially with step $i$ of the Lanczos method.

In this paper, we propose an alternative implementation of the Lanczos method, called the NQS Lanczos method, which consists of two parts: the supervised-learning Lanczos (SLL) algorithm and the VMC optimization part. The SLL algorithm further consists of two parts: supervised training and diagonalization. Through supervised training, the SLL algorithm represents Lanczos states with NQSs, thereby avoiding the calculation of the expectations $\langle H^{2i + 1} \rangle$. When the supervised learning is terminated, a set of basis NQSs are constructed. Then a Hamiltonian matrix is constructed with these basis NQSs. Diagonalizing the Hamiltonian matrix, a superposition state is obtained. The VMC optimization part are used to optimize the amplitude part of the superposition state. When the VMC optimization part is terminated, an optimized superposition state and the improved energy are obtained. Using the optimized superposition state as a new state for the next iteration, the full loop of the NQS Lanczos method is achieved by repeating the procedure.

We tested the NQS Lanczos method in high-frustration regions of the two-dimensional Heisenberg $J_1$-$J_2$ model ($ 0.5 \lesssim J_2/J_1 \lesssim 0.6$) on square lattices with linear size $L=4,6 $, and $10$. The results show that the NQS Lanczos method can significantly improve the energies of the systems.

The organization of this paper is as follows. In Section~\ref{sec:method}, we first introduce the structure of the NQSs used for supervised learning. Next, we show the details of the supervised learning, including the loss functions and the optimization strategy employed. Following this, we introduce a VMC optimization scheme specifically designed for the amplitude component. Finally, building on the aforementioned steps, we present the complete procedure of the NQS Lanczos method. In Section~\ref{sec:results}, we show the calculated results for the two-dimensional Heisenberg $J_1$-$J_2$ model and provide an analysis and discussion. Lastly, in Section~\ref{sec:summary}, we discuss the strengths and limitations of the NQS Lanczos method and provide an outlook for future improvements.

\begin{figure*}[t!]
  \includegraphics[width=16cm]{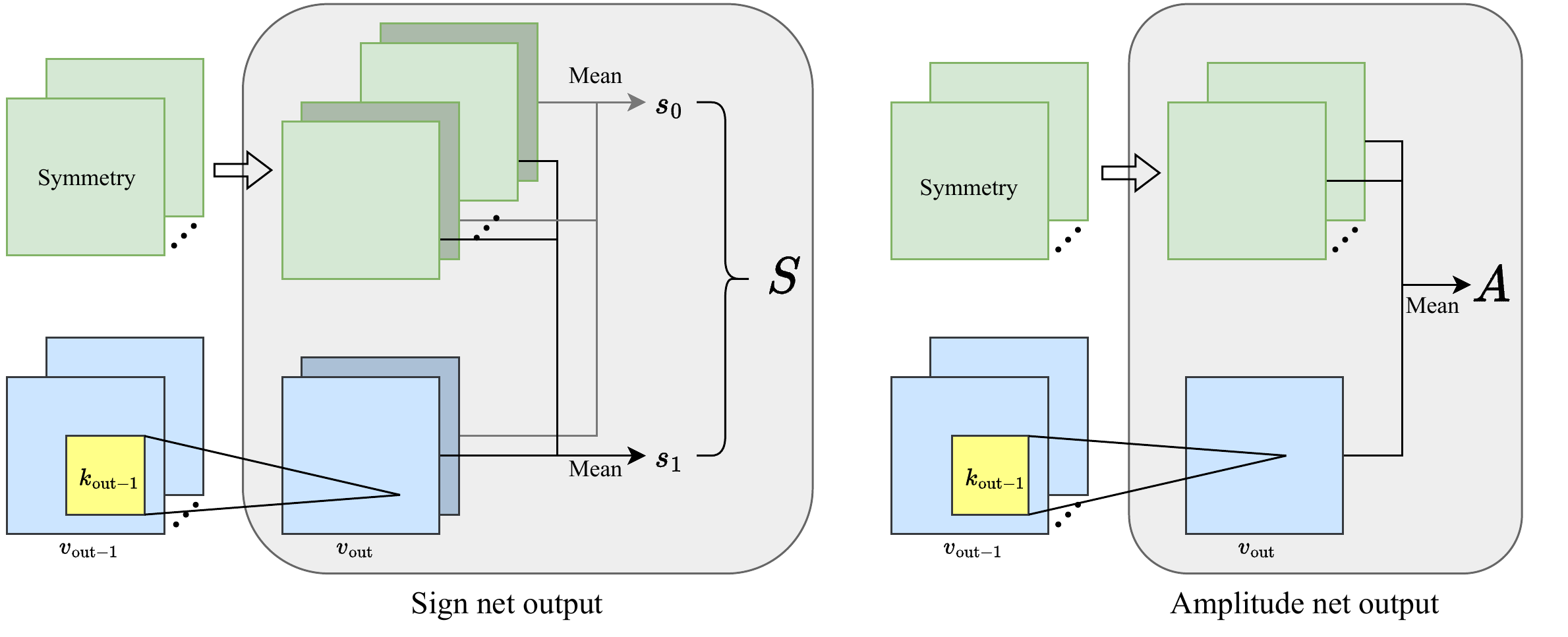}
  \caption{Schematic of the structure of the output layers of the sign network and the amplitude network. $v_{\text{out}}$ represents the output of the output layer, and $k_{\text{out}-1}$ denotes the convolutional kernel of the output layer. The flow labeled by symmetry means data argumentation by symmetry operations, which includes the spin-flip and the $\mathcal{C}_{\mathrm{4v}}$ symmetries. The output layer of the amplitude network has a single channel, while that of the sign network has two channels, which are used for the binary classification task of the sign (positive or negative). The outputs of the sign and amplitude network are denoted as $S$ and $A$, respectively. The corresponding wave function can be written as $\psi = S \cdot e^{A}$.}\label{fig:net}
\end{figure*}


\section{Method}
\label{sec:method}
In this section, we introduce a representation of the NQS. Based on this representation, we introduce the details of the SLL algorithm and design the loss function and optimization strategy. To address issues arising from underfitting in the SLL algorithm, we introduce a VMC optimization scheme to optimize the states further. The section concludes with an overview of the complete NQS Lanczos method.

\subsection{Neural-network quantum states}
One challenge in the optimization of NQS is the difficulty in handling the sign structure~\cite{Claudio2020_NQS_SignProblem,westerhout2023SignStructure}. In Ref.~\cite{Westerhout2020_difference_Sign_amp}, the optimization problems related to the sign and amplitude for frustrated systems were analyzed. Ref.~\cite{Jqwang2024aCNN} explored the case where only the amplitude is optimized, using a fixed sign structure. Choosing an appropriate neural-network architecture will facilitate the optimization. Various network architectures, such as RBM~\cite{Salakhutdinov2009DeepRBM}, multilayer perceptron~\cite{Rumelhart1986MLP}, recurrent neural network~\cite{Alex2013DeepRNN}, CNN~\cite{1Lecun1998CNN_LeNet}, and vision transformer~\cite{Dosovitskiy2021ViT}, have been used to represent NQSs~\cite{Lange2024NQSReview}. Among these, CNNs have been favored, with their intrinsic translation equivariance. Here we choose to use the real-valued CNN used in Ref.~\cite{Jqwang2024aCNN}, denoted as aCNN. The network is divided into two parts: the sign network ($\text{SNet}$) and the amplitude network ($\text{ANet}$).

The aCNN employs deep CNNs with shortcut connections as its main architecture. The shortcut connection is described in Ref.~\cite{he2015ResnetV1} (referred to as ResNet-v1). We have modified it to incorporate the structure presented in Ref.~\cite{he2016ResnetV2} (referred to as ResNet-v2). It has been widely demonstrated in the field of machine learning that ResNet-v2 is superior to ResNet-v1 in terms of network stability and expressiveness. The input to both the sign network and the amplitude network is a spin configuration $\sigma$ of the system. The only difference between the sign network and the amplitude network lies in the output layer, as shown in Fig.~\ref{fig:net}. 
The output layer of the amplitude network has only one channel, while the sign network has two. By calculating the mean value of the elements within the output channel of the amplitude network, the amplitude output is obtained and denoted as $A$. Meanwhile, for the sign network, the mean of each channel is computed to obtain $s_0$ and $s_1$, which are used for the binary classification task. The output of the sign network is denoted as $S$. Together, the sign and amplitude networks form a representation of a wave function. This representation is used in the following sections to represent the wave function of the Lanczos state at step $i$, and the wave function is denoted as $\psi_i = S_i \cdot e^{A_i}$.

Additionally, the implementation of lattice symmetries is divided into two parts. Translation symmetry is easily realized by maintaining a constant hidden layer size and averaging over the output layer. Spatial rotation, mirror reflection, and spin-flip symmetries are realized by performing data augmentation on the input configurations and averaging the outputs. This corresponds to the flow labeled by symmetry in Fig.~\ref{fig:net}.

\subsection{The Lanczos algorithm}
The Lanczos method~\cite{Lanczos1952Method} was first introduced by Cornelius Lanczos in 1952. In 2013, Hu \textit{et al.}~\cite{Sorella2013_VMC} used this approach to further improve the states obtained by the VMC method in the simulation of many-body systems, achieving promising results at that time. The procedure of the Lanczos algorithm is as follows.

i. Give an initial state $|\psi_0\rangle$. This state, which is not orthogonal to the ground state, can be arbitrary.

ii. Define $|v_1 \rangle = H | \psi_0 \rangle - a_0 | \psi_0 \rangle$, where $a_0 = {\langle \psi_0 | H | \psi_0 \rangle} / {\langle \psi_0 | \psi_0\rangle}$ is the expectation of the Hamiltonian on the state $|\psi_0\rangle$. Let $ b_1 |\psi_1 \rangle = |v_1 \rangle $, where $b_1 = \sqrt{{\langle v_1 |v_1 \rangle} / {\langle \psi_0 | \psi_0 \rangle}}$ is the normalization factor that normalizes $|v_1 \rangle$ with respect to $|\psi_0 \rangle$, namely $\langle \psi_1 | \psi_1 \rangle  = \langle \psi_0|\psi_0 \rangle$.

iii. Define 
\begin{align}
    |v_{i+1} \rangle = H | \psi_{i} \rangle - a_{i} | \psi_{i} \rangle - b_{i} | \psi_{i-1} \rangle,
    \label{eq:v_i}
\end{align}
where
\begin{align}
    a_i &= \frac{\langle\psi_i|H|\psi_i\rangle}{\langle\psi_i|\psi_i\rangle}.
    \label{eq:a}
\end{align}
The Lanczos state $| \psi_{i+1} \rangle$ is constructed as 
\begin{align}
    |\psi_{i+1} \rangle &=  \frac{|v_{i+1} \rangle} {b_{i+1}},
    \label{eq:psi_i}
\end{align} where
\begin{align} 
    b_{i+1} = \sqrt{\frac{\langle v_{i+1} |v_{i+1} \rangle}{\langle \psi_0|\psi_0\rangle}}.
    \label{eq:b}
\end{align}
Step iii can be used for all subsequent iterations.

iv. After $p$ iterations, a set of $p + 1$ Lanczos vectors $|\psi_i\rangle$, $i = 0,\cdots,p$, can be obtained, which forms an orthonormal basis set spanning a reduced Hilbert space. With this basis the Hamiltonian is represented as a tridiagonal matrix,

    \begin{align}
        T =
        \begin{bmatrix}
            a_0    & b_1    & 0      & \cdots & 0      \\
            b_1    & a_1    & b_2    & \cdots & 0      \\
            0      & b_2    & a_2    & \cdots & 0      \\
            \vdots & \vdots & \vdots & \ddots & \vdots \\
            0      & 0      & 0      & \cdots & a_p \\
        \end{bmatrix}. \nonumber
    \end{align}

v. Diagonalize $T$ to obtain eigenvalues and corresponding eigenvectors.

\subsection{The supervised-learning Lanczos algorithm}
In this work, we use the Lanczos state $|\psi_i\rangle$ at Lanczos step $i$ as the target of supervised learning. Through supervised learning, the sign network and amplitude network of the NQS are optimized so that the NQS can be used as an approximate representation of $|\psi_i\rangle$. The NQS and supervised learning are combined to implement the Lanczos procedure.

First, we use aCNN as the initial state, denoted as $|\psi^{\text{net}}_0\rangle$. Then the Lanczos state $|\psi_i \rangle$ at step $i$ is set as the target state, denoted as $|\psi^\mathrm{trg}_i\rangle$, to be learned. Through Monte Carlo sampling, the coefficients $a_i$ and $b_{i + 1}$ can be calculated. Then the target state $|\psi^\mathrm{trg}_i\rangle$ can be expressed with these coefficients and the previrous states.

Take a configuration $\sigma$ as a training sample for supervised learning. The sign and amplitude of $\psi^\mathrm{trg}_i(\sigma)$ are taken as labels for the supervised learning of the sign network and the amplitude network, respectively. Then, the sign network $\text{SNet}_i$ and the amplitude network $\text{ANet}_i$ can be trained through supervised learning. Once the losses converge, the wave function $\psi^{\text{net}}_i = S_i \cdot e^{A_i}$ of the Lanczos state at step $i$ is obtained, that is, the Lanczos state is represented by a NQS.

This procedure completes the iteration of the supervised learning part of the SLL algorithm, which generates Lanczos states represented by NQSs.

When the iteration of the supervised training is completed, a set of basis NQSs will be obtained. Then a Hamiltonian matrix can be constructed with these NQSs. By diagonalizing the matrix, the lowest eigenvalue (i.e., the improved energy $E$) and the corresponding superposition state
\begin{align}
    |\Psi\rangle = \sum_{i=0}^{p} c_i |\psi_i^\mathrm{net} \rangle
\end{align}
can be obtained, where $c_i$ is the superposition coefficient. Details of the diagonalization process are given in Appendix~\ref{sec:diagonalization}.

\subsection{Loss function and optimization}
The effectiveness of using a NQS to approximate a Lanczos state depends on the performance of the supervised learning.

The goal of supervised learning of the amplitude network is to make the distribution described by $|\psi^\mathrm{net}_i|^2$ consistent with the distribution described by $|\psi^\mathrm{trg}_i|^2$. In deep learning field, the Kullback-Leibler (KL) divergence is often used to measure the difference between two distributions. However, in our tests, the KL divergence loss function performed poorly. We choose to use the loss function based on the mean squared error, $\mathbb{E}_{x \sim p}||y_x - f(x)||$. It is to minimize the difference squared $\left( |\psi^{\text{trg}}_{i}| - |\psi^{\text{net}}_{i}| \right)^2$ between the target and the prediction. To achieve this, samples should be obtained from the distributions described by both $ |\psi^{\text{trg}}_{i}|^2$ and $|\psi^{\text{net}}_{i}|^2$. The loss function of the amplitude part is
\begin{align}
    L_{\text{amp}} &= \frac{1}{2} \sum_{\sigma} p^{\text{trg}}_{i}(\sigma) \left(| \psi^{\text{trg}}_{i}(\sigma)| - |\psi^{\text{net}}_{i}(\sigma, \theta)| \right)^2 \nonumber \\ &+ \frac{1}{2} \sum_{\sigma} p^{\text{net}}_{i}(\sigma, \theta)  \left(| \psi^{\text{trg}}_{i}(\sigma)| - |\psi^{\text{net}}_{i}(\sigma, \theta)| \right)^2,
    \label{eq:amp_loss1}
\end{align}
where the probability distributions are
\begin{align}
    p^{\text{trg}}_{i}(\sigma) = \frac{\left| \psi^{\text{trg}}_{i}(\sigma) \right|^2}{\sum_{\sigma} \left| \psi^{\text{trg}}_{i}(\sigma) \right|^2}
\end{align}
and
\begin{align}
    p^{\text{net}}_{i}(\sigma, \theta) = \frac{\left| \psi^{\text{net}}_{i}(\sigma, \theta) \right|^2}{\sum_{\sigma} \left| \psi^{\text{net}}_{i}(\sigma, \theta) \right|^2}.
\end{align}

It is worth noting that the probability distribution $p^{\text{trg}}_{i}(\sigma)$ and the label $\psi^{\text{trg}}_{i}(\sigma)$ are both determined by the previous NQSs, whose parameters are fixed and are not involved in optimization here. While the probability distribution $p^{\text{net}}_{i}(\sigma, \theta)$ contains the parameters $\theta$ to be optimized. The gradient with respect to these parameters must be considered.
According to the theory of automatic differentiation Monte Carlo (ADMC)~\cite{zhang2023ADMC}, the expectation of an observable $O(\sigma, \theta)$ is given by
\begin{align}
    \langle O(\sigma, \theta)\rangle_{p(\sigma, \theta)} = \frac{\langle \frac{p(\sigma, \theta)}{\perp(p(\sigma, \theta))} O(\sigma, \theta) \rangle_{\perp(p(\sigma, \theta))}}{\langle \frac{p(\sigma, \theta)}{\perp(p(\sigma, \theta))}\rangle_{\perp(p(\sigma, \theta))}},
    \label{eq:admc}
\end{align}
where $\perp(x)$ is the detach function, which was first introduced in Section III A of Ref.~\cite{zhang2023ADMC}, features $\perp(x) = x$ in forward propagation and ${\partial\perp(x)} / {\partial x} = 0$ in backward propagation. $x$ does not propagate gradients in the automatic differentiation process. This reformulation applies to both normalized and unnormalized probability distributions. The loss function for the amplitude network can be rewritten as
\begin{align}
    L_{\text{amp}} &= \frac{1}{2} \mathbb{E}_{\sigma\sim p^{\text{trg}}_{i}} \left(| \psi^{\text{trg}}_{i}(\sigma)| - |\psi^{\text{net}}_{i}(\sigma, \theta)| \right)^2 \nonumber \\
    &+ \frac{1}{2} \frac{\mathbb{E}_{\sigma\sim p^{\text{net}}_{i}}  \frac{p^{\text{net}}_{i}(\sigma, \theta)}{\perp(p^{\text{net}}_{i}(\sigma, \theta))}  \left(| \psi^{\text{trg}}_{i}(\sigma)| - |\psi^{\text{net}}_{i}(\sigma, \theta)| \right)^2 }  {\mathbb{E}_{\sigma\sim p^{\text{net}}_{i}} \frac{p^{\text{net}}_{i}(\sigma, \theta)}{\perp(p^{\text{net}}_{i}(\sigma, \theta))}}.
    \label{eq:amp_loss}
\end{align}

\begin{figure}[ht]
  \centering 
  \includegraphics[width=\columnwidth]{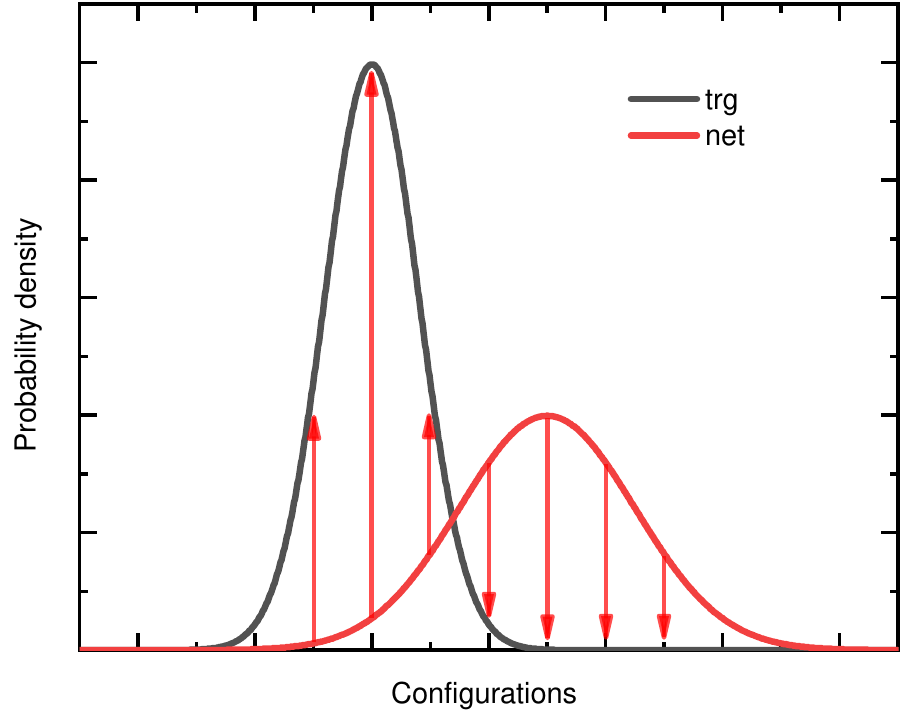}
  \caption{Illustration of the loss function of the amplitude network. The horizontal axis represents the configurations, while the vertical axis represents the probability density. The curves labeled by `trg' and `net' serve as schematic representations of the distributions described by $|\psi^{\text{trg}}_{i}(\sigma)|^2$ and $|\psi^{\text{net}}_{i}(\sigma, \theta)|^2$, respectively. The arrows indicate the expected changes in the `net' distribution when supervised learning. Both those two distributions are used to generate the samples for supervised learning.}
  \label{fig:amp_loss}
\end{figure}

The illustrations of this loss function are shown in Fig.~\ref{fig:amp_loss}. The horizontal axis represents the configurations, while the vertical axis represents the probability density. The arrows indicate the expected changes in the ‘net’ distribution when supervised learning. Sampling from these two distributions separately ensures that $| \psi^{\text{net}}_{i}(\sigma)|$ can be adjusted for the configurations $\sigma$ where $| \psi^{\text{trg}}_{i}(\sigma)|$ are extremely small.

The loss defined in Eq.~\eqref{eq:amp_loss1} is the MSE loss. Another choice is to adopt the fidelity between the quantum states as the loss function, as employed in Refs.~\cite{Sinibaldi2023unbiasingtime,gravina2025neuralprojectedquantumdynamics,jonsson2018neuralnetworkstatesclassicalsimulation}. The loss function based on fidelity is given by
\begin{align}
    L(\psi^\mathrm{net}_{i}(\sigma, \theta), \psi^\mathrm{trg}_i(\sigma))
    &= - \log \left[ \frac{ \left| \langle \psi^\mathrm{net}_{i}(\sigma, \theta) | \psi^\mathrm{trg}_i(\sigma) \rangle \right| }{\left| { | \psi^\mathrm{net}_{i}(\sigma, \theta) \rangle } \right|  \left| { | \psi^\mathrm{trg}_i(\sigma) \rangle } \right| } \right] \nonumber \\
    &= - \log \sqrt{\alpha \beta },
    \label{eq:fidelity_loss}
\end{align}
where
\begin{align}
    \alpha &= \left| \left< \frac{\psi^\mathrm{trg}_i(\sigma)}{\psi^{\mathrm{net}}_{i}(\sigma, \theta) (\sigma)}\right>_{\psi^{\mathrm{net}}_{i}(\sigma, \theta)} \right|, \nonumber \\
    \beta &= \left| \left< \frac{\psi^{\mathrm{net}*}_{i}(\sigma, \theta) (\sigma)}{\psi^{\mathrm{trg}*}_i(\sigma)} \right>_{\psi^\mathrm{trg}_i(\sigma)} \right|.
    \label{eq:fidelity_loss_alpha}
\end{align}

The derivative of the loss function based on fidelity with respect to the parameter $\theta$ can be referred to Ref.~\cite{jonsson2018neuralnetworkstatesclassicalsimulation}. However, the optimization of this loss function is difficult due to its inherent numerical instability, which often leads to overflow during training. To address this, we have implemented two distinct strategies. Strategy 1 (denoted ad s1) introduces a regularization term to the loss function, constraining the amplitudes within an appropriate range. Strategy 2 (denoted as s2) involves dynamic monitoring during optimization. If the components $\alpha$ or $\beta$ in the loss function exhibit anomalous behavior, the corresponding term is detached during backpropagation and is temporarily excluded from optimization until its values return to a stable, reasonable range.

While these strategies involve explicit manual design, they often serve as effective auxiliary treatments that stabilize the optimization process in our tests. A comparative analysis of the convergence behavior for different loss functions is provided in Section~\ref{sec:results}.B. In subsequent numerical experiments, the MSE loss demonstrated robust and competitive performance. Consequently, unless stated otherwise, we choose the MSE loss as the default throughout the rest of this paper.

The optimization of the sign network is a binary classification task. The target is to learn the sign of $\psi^{\text{trg}}_{i}(\sigma)$. In this work, we choose to use the commonly used cross-entropy loss function
\begin{align}
    L_{\text{sign}} &= -\mathbb{E}_{\sigma} \left[ \hat{y}_{\sigma}\log q_\sigma + (1-\hat{y}_{\sigma})\log(1-q_\sigma) \right],
    \label{eq:sign_loss}
\end{align}
where $\hat{y}_{\sigma}$ denotes the label, and $q_\sigma$ represents the prediction. Unlike the amplitude part, training the sign network with samples from only the target distribution is sufficient.

In addition, sampling from the distribution described by $|\psi_i^\mathrm{trg}|^2$ to generate a fixed dataset before training is beneficial. The data in the dataset can be read directly and used for network optimization.

In practice, samples and labels from $|\psi_i^\mathrm{trg}\rangle$ are prepared in advance, denoted as $\sigma_{\text{trg}}$ and  $\psi^{\text{trg}}_i(\sigma_{\text{trg}})$. For the amplitude optimization, we sample $\sigma_{\text{net}}$ from the distribution described by $|\psi^\mathrm{net}_i|^2$, and calculate its label value $\psi^{\text{trg}}_i ( \sigma_{\text{net}} )$. Then both $\sigma_{\text{trg}}$ and $\sigma_{\text{net}}$ are passed into the network for forward propagation. For the sign optimization, samples are read directly from the dataset without the need for additional sampling.

\subsection{VMC on the superposition state}
In the tests of aCNN~\cite{Jqwang2024aCNN}, the energies achieved by optimizing only the amplitude part of the wave function are more accurate than those of the complex-valued CNN~\cite{Choo2019_CNN_ComplexValued} with the same number of parameters. This inspires us to further optimize the amplitude of the superposition of NQSs to mitigate the negative effects of insufficient convergence in the supervised learning. We select an amplitude network $\text{ANet}_j$ with $ j\neq0$ to do a VMC optimization, while fixing all the parameters of the other networks as well as the superposition coefficients.

During the VMC optimization, we use energy as the loss function and employ the ADMC method~\cite{zhang2023ADMC} to calculate the derivative of energy with respect to the network parameters. The popular Adam algorithm~\cite{Kingma2015Adam} is used to update the parameters. The energy is expressed as
\begin{align}
    E &= \left< H \right> = \langle E_\mathrm{loc}(\sigma, \theta_i)\rangle_{p(\sigma, \theta_i)} \nonumber \\
    &= \frac{\langle \frac{\Psi^2(\sigma, \theta_i)}{\perp(\Psi^2(\sigma, \theta_i))} E_\mathrm{loc}(\sigma, \theta_i) \rangle_{\perp({p(\sigma, \theta_i)})}}{\langle \frac{\Psi^2(\sigma, \theta_i)}{\perp(\Psi^2(\sigma, \theta_i))}\rangle_{\perp({p(\sigma, \theta_i)})}},
    \label{eq:admc_loss}
\end{align}
where $\theta_j$ represents the trainable parameters of the $j$-th amplitude network $\text{ANet}_j$,
\begin{align}
    E_\mathrm{loc}(\sigma, \theta_j) = \frac{\langle \sigma|H|\Psi_{\theta_j} \rangle}{\langle\sigma|\Psi_{\theta_j}\rangle} = \sum_{\sigma'}H_{\sigma\sigma'}\frac{\Psi(\sigma', \theta_j)}{\Psi(\sigma, \theta_j)}
\end{align}
is the local energy of the system,
\begin{align}
    p(\sigma, \theta_j)=\frac{|\Psi(\sigma, \theta_j)|^2}{\sum_{\sigma'}{|\Psi(\sigma', \theta_j)|^2}}
\end{align}
is the probability distribution, and $\langle \cdot \rangle$ represents the expectation.

The VMC optimization part effectively optimizes the amplitude network $\text{ANet}_j$ further. The goal is not to make $|\psi^\mathrm{net}_j|$ match $|\psi^\mathrm{trg}_j|$ but to make the overall superposition state $|\Psi\rangle$ approach the true ground state. After optimizing each $\text{ANet}_j$ with $0<j \leq p$ (note that $\text{ANet}_0$ has already been obtained through amplitude optimization and does not need further optimization), we obtain a new state $|\tilde{\Psi}\rangle$. This new state can then be used as the updated $|\psi^\mathrm{net}_0\rangle$ for the next NQS Lanczos loop to further improve the results.

\subsection{The NQS Lanczos method}

\begin{figure*}[t!]
  \includegraphics[width=16cm]{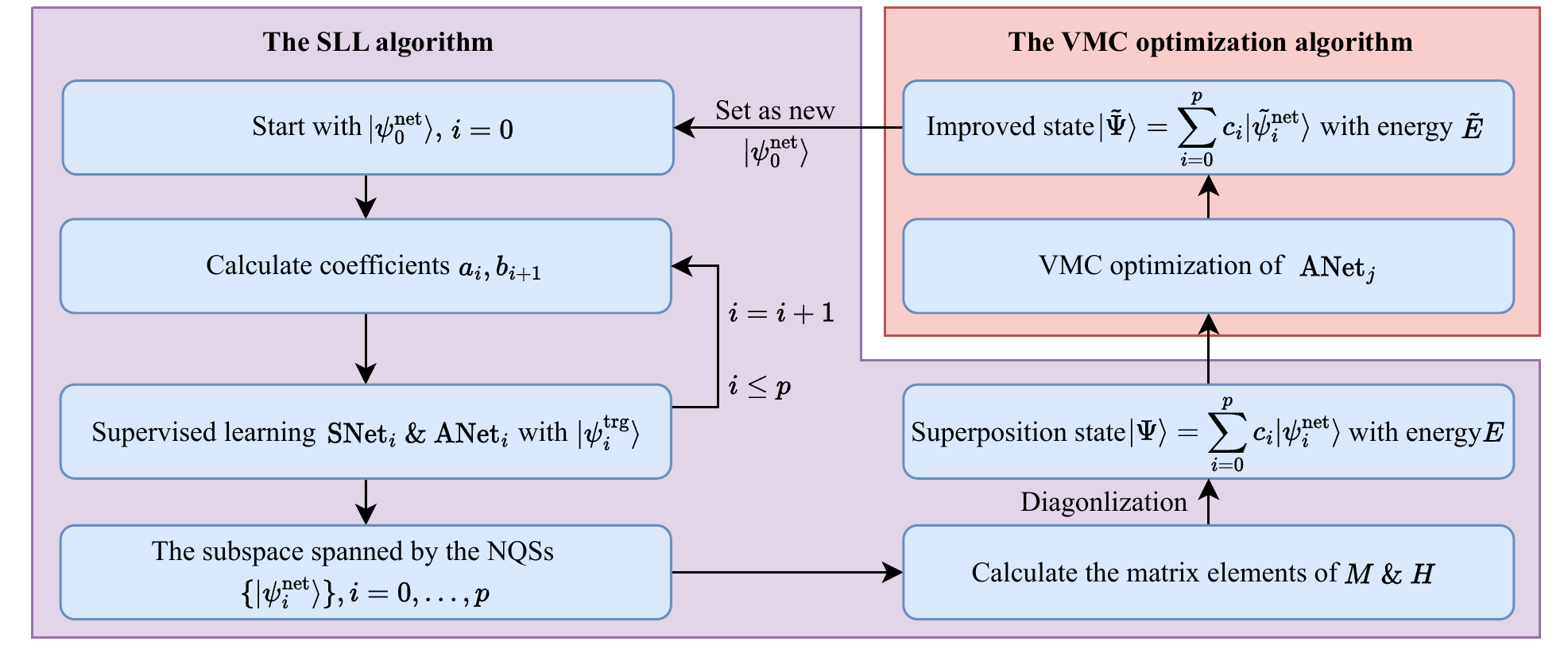}
  \caption{Flowchart of the NQS Lanczos method. The method begins with $|\psi^\mathrm{net}_0\rangle$ as a starting point and ultimately outputs the improved energy $\tilde{E}$ and state $|\tilde{\Psi}\rangle$. $|\tilde{\Psi}\rangle$ can be used as a new $|\psi^\mathrm{net}_0\rangle$ for the next NQS Lanczos loop. The NQS Lanczos method consists of two parts, the SLL algorithm and the VMC optimization part. The SLL algorithm further consists of two parts: supervised learning of Lanczos states and diagonalization of the Hamiltonian matrix. The output of the SLL algorithm consists of the superposition state $|\Psi\rangle$ and the improved energy $E$. The VMC optimization part further optimizes $|\Psi\rangle$ and gives the final output of the NQS Lanczos method. The number of the Lanczos steps is denoted as $p$. The matrices $M$ and $H$ are used to produce the superposition state, as shown in Appendix~\ref{sec:diagonalization}.}
  \label{fig:SLL_algorithm}
\end{figure*}

In this section, we outline the entire procedure of the NQS Lanczos method, as shown in Fig.~\ref{fig:SLL_algorithm}. The NQS Lanczos method consists of two parts, the SLL algorithm and the VMC optimization part. The first step is to provide an initial state $|\psi^\mathrm{net}_0\rangle$, based on which the coefficients $a_i$ and $b_{i+1}$ are calculated. Then the target $|\psi^\mathrm{trg}_i\rangle$ is constructed, according to which, samples and labels are generated for supervised learning. Through the supervised learning, both the sign and the amplitude networks are optimized. The wave function $\psi^\mathrm{net}_i = S_i e^{A_i}$ represented by NQS is obtained. As the Lanczos step $i$ goes, a set $\{ |\psi^\mathrm{net}_i\rangle \}$ ($i=0, \dots ,p$) is obtained.

Next, we construct a Hamiltonian matrix $H$ and a Hermitian matrix $M$ (see Appendix~\ref{sec:diagonalization}) in the space spanned by the set of the NQSs. Diagonalizing the Hamiltonian matrix, a superposition state $|\Psi\rangle$ composed of the NQSs and an improved energy $E$ are obtained.

Finally, VMC is applied to optimize all amplitude networks $\text{ANet}_j$ (except $\text{ANet}_0$) of the superposition state $|\Psi\rangle$. This produces the final improved state $|\tilde{\Psi}\rangle$, on which the Hamiltonian expectation is calculated to obtain the final improved energy $\tilde{E}$.

The superposition state $|\tilde{\Psi}\rangle$ can be used as a new initial state $|\psi^\mathrm{net}_0\rangle$. Then the NQS Lancozs method can be repeated to further improve the results.

Looking at the overall flowchart of the NQS Lanczos method, one can see that the computational complexity grows linearly with the number of the Lanczos steps $p$. The largest computational cost comes from the calculation of the parameters $a_i$ (the expectation of $H$), the parameters $b_i$ (involving $H|\psi^\mathrm{net}_{i}\rangle$), and the matrix elements of the matrices $M$ and $H$. The NQS Lanczos method avoids the calculation of the expectation $\langle H^{2i+1} \rangle$ that are required in Ref.~\cite{chen2022lanczos}. The growth rate of the computational cost in Ref.~\cite{chen2022lanczos} follows $\left( 4N^2 \right)^{2i+1}$, where $N$ is the number of lattice sites. This limits the maximum number of the Lanczos steps $p$. The computational cost of the NQS Lanczos method enables more Lanczos steps to be performed even with limited computational resources.

\section{Numerical results and discussions}
\label{sec:results}
In this section, we apply the NQS Lanczos method to the two-dimensional Heisenberg $J_1$-$J_2$ model on the square lattices with $L = 4, 6$, and $10$. The Hamiltonian is
\begin{equation}
    H = J_1 \sum_{\langle ij \rangle}{\hat{S_i} \cdot \hat{S_j}} + J_2    \sum_{\langle \langle ij \rangle \rangle}{\hat{S_i} \cdot \hat{S_j}},
\end{equation}
where $J_1$ and $J_2$ are the coupling strengths of nearest-neighbor sites and the next nearest-neighbor sites (denoted as $\langle ij \rangle$ and $\langle \langle ij \rangle \rangle$), respectively. This model is strongly frustrated when $J_2/J_1 $ is around $0.5$. It is challenging to theoretically study this region. As the ratio $J_2 / J_1$ increases from $0$, the system undergoes a phase transition~\cite{Zheng-ChengGu2022_TensorNetwork+PEPS} from a Neel antiferromagnet to a gapless quantum spin liquid, then to a valence-bond solid, and finally to a collinear antiferromagnet. In this section, we test the NQS Lanczos method in the highly frustrated region. The test results from different stages (SLL and VMC) of the method are presented to comprehensively demonstrate the effectiveness of the method.

To avoid confusion with traditional VMC, we introduce the term ``VMC Lanczos" (denoted as VMCL) to specifically label the results obtained from performing VMC on the superposition state.

In each test, $|\psi^\mathrm{net}_0\rangle$ is obtained through the optimization of the amplitude network with a sign structure, which is fixed to the Marshall sign rule (MSR)~\cite{1955MarshallSign, Jqwang2024aCNN}. The following tests all use the checkerboard patterned sign structure. Additionally, due to the highly rugged landscape of the wave function, the optimization results can be influenced by several factors, such as the initialization method of the network parameters, the seed of the random numbers for the initialization, the learning rate, and the optimization algorithm. The networks to be trained in the supervised learning are initialized by Kaiming initialization~\cite{he2015KaimingInitilization}. To reduce the fluctuation caused by initialization, we use four different seeds to initialize the network and retain the result with the best optimization performance. The Adam algorithm is used for all optimization. All calculations in this work are performed on a single NVIDIA 3090 GPU.

\subsection{Test of the supervised learning Lanczos algorithm on the $4 \times 4$ lattice}
\begin{figure}[ht]
  \centering 
  \includegraphics[width=7cm]{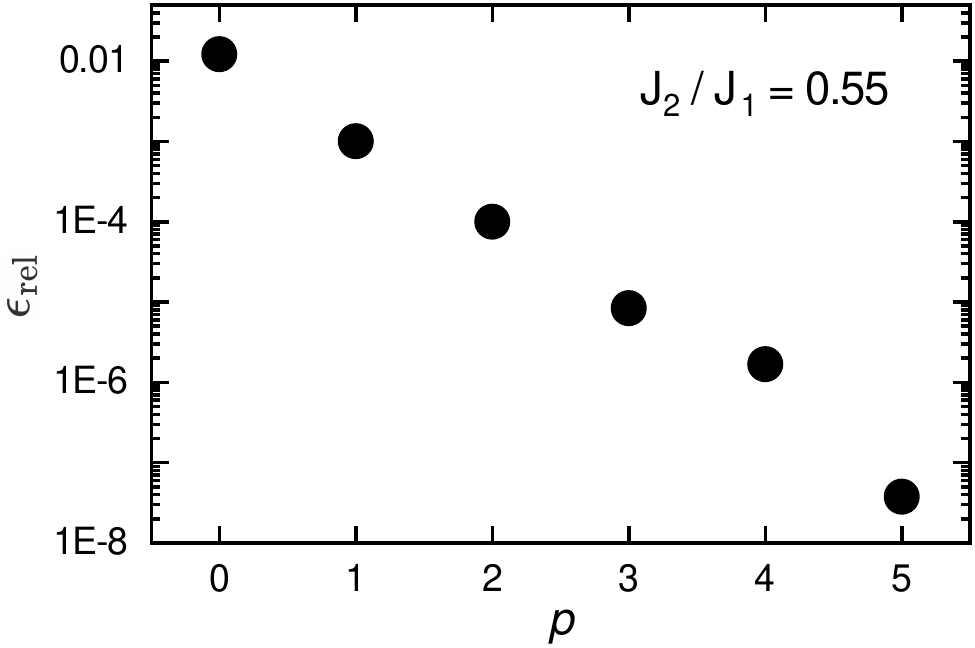}
  \caption{Improved energies on the $4 \times 4$ lattice. The horizontal axis represents the number of Lanczos steps. The vertical axis indicates the relative error of the improved energies with respect to the energy obtained from ED, $\epsilon_\text{rel} = (E - E_\text{ED}) / | E_\text{ED}|$.} As the number of the Lanczos steps increases, the energy rapidly converges to the exact one.
  \label{fig:L4}
\end{figure}

We first test our method on the $L = 4$ square lattice with $J_2 / J_1 = 0.55$. The test results are shown in Fig.~\ref{fig:L4}. The improved energies are obtained from the SLL algorithm. The horizontal axis represents the number of the Lanczos steps. The vertical axis indicates the relative error of the improved energies with respect to the energy obtained from the exact diagonalization (ED) method, $\epsilon_\text{rel} = (E - E_\text{ED}) / | E_\text{ED}|$. The relative error decreases exponentially from $ 1.20 \times 10^{-2}$ at $p = 0$ to $3.75\times10^{-8}$ at $p = 5$.

For the $L = 4$ square lattice, the Hilbert space can be traversed easily, the coefficients $a_i$ and $b_{i+1}$ can be calculated accurately, and the loss can be optimized to $10^{-8}$. In this case, the supervised trained states are almost strictly orthogonal. The combination of the Lanczos method, supervised learning, and NQSs is successful. The SLL algorithm can perfectly implement the Lanczos method without the need for the VMC optimization part.

\subsection{Test of the supervised-learning Lanczos algorithm on the $6 \times 6$ and $10 \times 10$ lattices}
In the test of the $4\times4$ lattice, the supervised learning can effectively capture all the characteristics of the target. This enables the Lanczos method to be implemented perfectly with NQSs. However, as the lattice size increases, the Hilbert space grows exponentially and the configurations cannot be traversed. The dataset cannot describe a complete picture of the wave function. In this situation, capturing the main characteristics of the target to achieve an effective representation of the target state poses a significant challenge to the learning and generalization abilities of the neural network.

\begin{figure}[t!]
  \includegraphics[width=\columnwidth]{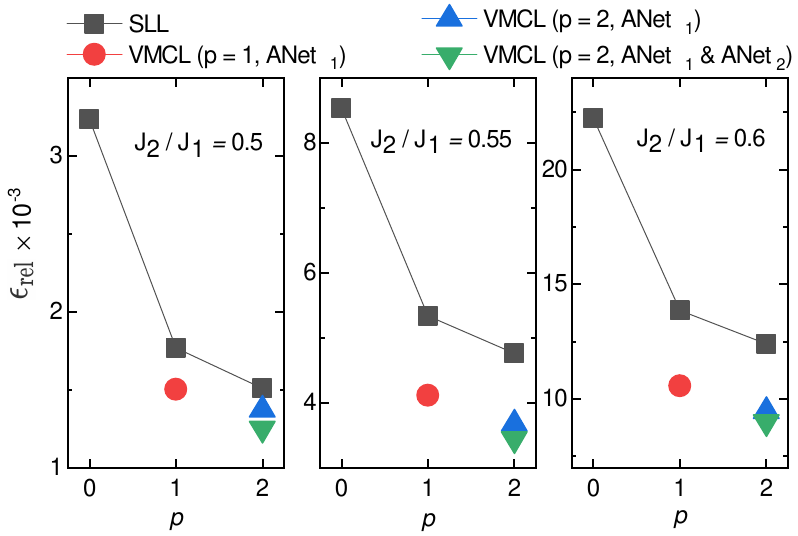}
  \caption{Improved energies on the $6 \times 6$ lattice. The model is test with $J_2/J_1 = 0.5$, $0.55$, and $0.6$. The vertical axis indicates the rescaled relative error, $\epsilon_\text{rel} \times 10^{-3}$, of the improved energies with respect to the energy obtained from ED. At $p \neq 0$, the points of SLL indicate the energies obtained by the SLL algorithm. At $p = 0$, the points of SLL indicate the energies of the initial states. The points of VMCL ($p = n$, $\text{ANet}_j$) indicate the energies obtained by the VMC. The superposition state consists of $n+1$ NQSs. $\text{ANet}_j$ indicates that the amplitude network of Lanczos step $j$ is optimized in the VMC.}
  \label{fig:L6}
\end{figure}

\begin{figure}[t!]
  \includegraphics[width=\columnwidth]{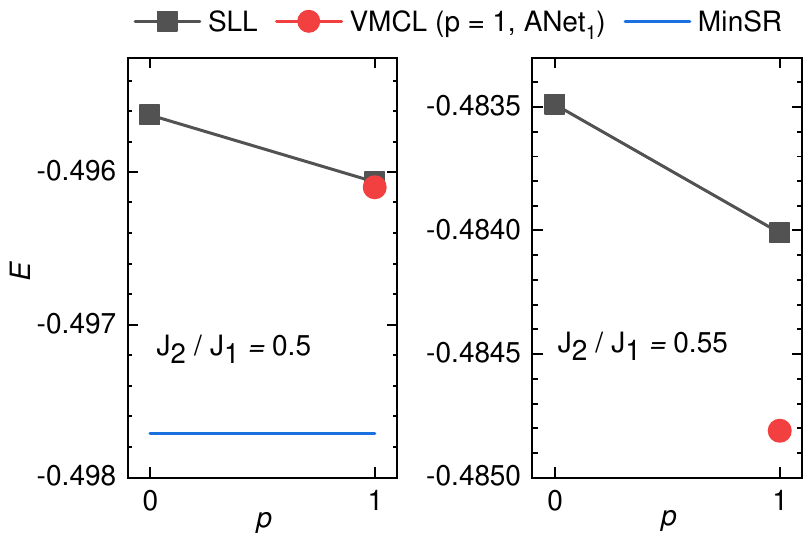}
  \caption{Improved energies. Similar to Fig.~\ref{fig:L6}, but on the $10 \times 10$ lattice. The result of MinSR is from Ref.~\cite{Chen2024MinSR}.}
  \label{fig:L10}
\end{figure}

The performance of the SLL algorithm on the square lattices with $L = 6$ and $L = 10$ is tested in this section. For the $L = 6$ lattice, the aCNN is used with the ResNet-v2 connection. We test the cases of $p = 1$ and $p = 2$, and the results are shown in Fig.~\ref{fig:L6}. The vertical axis indicates the rescaled relative error, $\epsilon_\text{rel} \times 10^{-3}$, of the improved energies with respect to the energy obtained from ED. For $L = 10$, we continue to use the aCNN with the ResNet-v1 connection and test the case of $p = 1$, and the results are shown in Fig.~\ref{fig:L10}. The points with $p = 0$ indicate the energies of the initial Lanczos states, while for $p > 0$, the points labeled by SLL indicate the improved energies obtained by the SLL algorithm.

As shown in Figs.~\ref{fig:L6} and~\ref{fig:L10}, compared to $p = 0$, the energies obtained by the SLL algorithm are significantly improved at $p = 1$ for both $L = 6$ and $L = 10$. Furthermore, the improvement remains highly effective at $p = 2$. This shows that the SLL algorithm remains effective even on lattices with larger sizes.

\begin{table}[t!]
  \begin{center}
    \caption{Accuracy of sign prediction of the sign net and loss of the amplitude net.}
    \label{tab:accuracy_loss}
    \setlength{\tabcolsep}{1.5mm}{
      \begin{tabular}{l l r@{.}l r@{.}l r@{.}l r@{.}l r@{.}l r@{.}l}
        \hline
        \hline
        \multicolumn{1}{l}{} & \multicolumn{1}{l}{} & \multicolumn{6}{c}{Sign accuracy} & \multicolumn{6}{c}{Amplitude loss} \\
        \hline
        \multicolumn{1}{l}{} & \multicolumn{1}{l}{$J_2/J_1$} & \multicolumn{2}{l}{$0.5$} & \multicolumn{2}{l}{$0.55$} & \multicolumn{2}{l}{$0.6$} & \multicolumn{2}{l}{$0.5$} & \multicolumn{2}{l}{$0.55$} & \multicolumn{2}{l}{$0.6$} \\
        \hline
        $L=6$  & $|\psi^\mathrm{net}_1\rangle$ & 95&8\%   & 96&4\%   & 98&8\%      & 0&21    & 0&15     & 0&10         \\
               & $|\psi^\mathrm{net}_2\rangle$ & 86&0\%   & 87&0\%   & 87&0\%      & 0&44    & 0&44     & 0&43          \\
        \hline
        $L=10$ & $|\psi^\mathrm{net}_1\rangle$ & 91&2\%   & 89&6\%   & \multicolumn{2}{c}{} & 0&46  & 0&32  &  \multicolumn{2}{c}{} \\
        \hline
        \hline
      \end{tabular}
    }
  \end{center}
\end{table}

\begin{figure*}[t]
  \includegraphics[width=16cm]{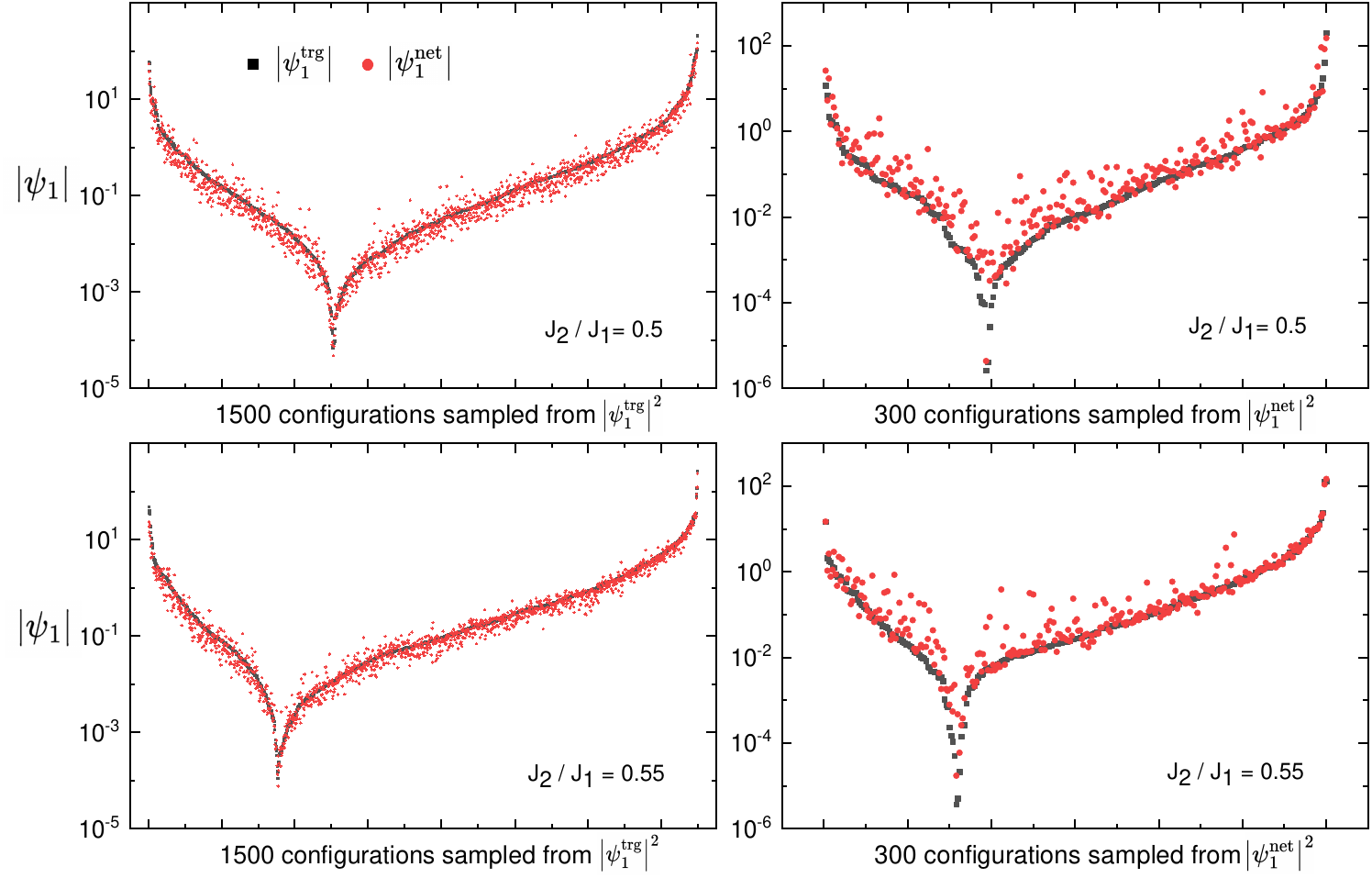}
  \caption{Target labels $\psi^\mathrm{trg}_1(\sigma)$ vs. predicted values $\psi^\mathrm{net}_1(\sigma)$. The test is on a square lattice with $L = 10$, $J_2/J_1 = 0.5$ and $0.55$. The horizontal axis represents the configurations sampled from the distributions described by $|\psi^{\text{trg}}_{1}|^2$ and $|\psi^{\text{net}}_{1}|^2$. These configurations are sorted by their labels in ascending order. The vertical axis represents the absolute values of $\psi^\mathrm{trg}_1(\sigma)$ and $\psi^\mathrm{net}_1(\sigma)$.}
  \label{fig:amp_lable_net}
\end{figure*}

In fact, the convergence of the loss functions in this test is not ideal. In Table~\ref{tab:accuracy_loss}, we present the prediction accuracy of the sign parts and the loss of the amplitude parts in the supervised learning. Compared to the $L = 4$ case, as the lattice size increases, the loss becomes much more difficult to converge. The accuracy of the sign prediction drops from 100\% for $L = 4$ to 95\% for $L = 6$ and to 89\% for $L = 10$. The order of magnitude of the amplitude loss increases rapidly from $10^{-8}$ for $L=4$ to $10^{-1}$ for $L=6$ and $10$.

Figure~\ref{fig:amp_lable_net} shows the comparison between the predicted values $\psi^\mathrm{net}_1(\sigma)$ and the labels $\psi^\mathrm{trg}_1(\sigma)$ in the test of the $L = 10$ lattice with $J_2/J_1 = 0.5$, and $0.55$. The horizontal axis represents the configurations. These configurations are sampled from the distributions described by $|\psi^{\text{trg}}_{1}|^2$ and $|\psi^{\text{net}}_{1}|^2$, respectively, and are sorted by their labels in ascending order. The vertical axis represents the absolute values of $\psi^\mathrm{trg}_1(\sigma)$ and $\psi^\mathrm{net}_1(\sigma)$. The predictions of the amplitude network are generally close to the labels, which means the main characteristics of the target are captured. However, the poor accuracy of the predictions indicates underfitting.

\begin{table}[b!]
  \begin{center}
    \caption{Comparison of amplitude overlap for different loss functions on the $L=6$ square lattice.}
    \label{tab:fidelity_amplitude}
    \setlength{\tabcolsep}{6.0mm}{
      \begin{tabular}{l r@{.}l r@{.}l }
        \hline
        \hline
        \multicolumn{1}{l}{Loss} & \multicolumn{2}{c}{$J_2/J_1 = 0.5$} & \multicolumn{2}{c}{$J_2/J_1 = 0.6$} \\
        \hline
        MSE                & 0&84626(3)    & 0&92769(7)          \\
        Fidelity s1        & 0&82627(4)    & 0&91683(6)        \\
        Fidelity s2        & 0&84742(3)    & 0&91936(5)          \\
        \hline
        \hline
      \end{tabular}
    }
  \end{center}
\end{table}

To identify the most effective loss function for supervised learning, we compared their convergence behaviors. For each strategy, we tuned key hyperparameters, including the learning rate and its decay schedule. The results from eight random seeds were evaluated, with the best-performing instances detailed in Table~\ref{tab:fidelity_amplitude}. This table reports the overlap between the optimized neural network wave function and the target wave function ($p=1$) for the $L=6$ square lattice. The MSE loss demonstrates relatively superior performance across various reference points and benefits from a simple functional form, establishing it as the optimal choice for this study.

Exploring more effective optimization strategies for loss function based on fidelity or more suitable forms of the loss function remains highly valuable for future algorithmic improvements. For example, the $L^2$ distance discussed in Ref.~\cite{sinibaldi2025timedependentneuralgalerkinmethod} presents an alternative.

As the lattice size increases, the supervised learning fails to optimize the loss function to a sufficient level of accuracy. This results in discrepancies between $|\psi^\mathrm{net}_i|$ and the target $|\psi^\mathrm{trg}_i|$.
In the test for $L = 10$, the training sets contain 1.2 million (approximately $2^{20}$) samples obtained by Monte Carlo sampling. This is minuscule compared to the Hilbert space dimension $2^{100}$. However, the supervised trained networks are still able to accurately predict more than 85\% of the signs and learn the key characteristics of the amplitudes. The generalization ability of the neural networks far exceeds our expectations.

It is foreseeable that as the state approaches the true ground state, the difficulty of the supervised learning task will increase. As the state gradually approaches the eigenstate, the difference between the two states $|\psi\rangle$ and $H|\psi\rangle$ will gradually decrease and become very small. The target of the supervised learning becomes a lot of small quantities mixed with a large number of small and insignificant values caused by computational errors. These small quantities can not be effectively identified and learned by the supervised learning. This problem will become even more severe when the Hilbert space cannot be traversed. We speculate that the reason why the SLL algorithm remains effective in the aforementioned tests even though the accuracy of the prediction of signs does not reach 100\% is that the supervised learning passively abandons the learning of useless and random signs associated with the labels.

The numbers of the parameters of the sign network and the amplitude network are 6,614 and 6,538, respectively, in the test on the 6 × 6 and 10 × 10 lattices. If computational resources are sufficient, using neural networks with more parameters and more samples can further reduce the loss function of supervised learning and yield better results.

\begin{figure}[ht!]
  \includegraphics[width=7cm]{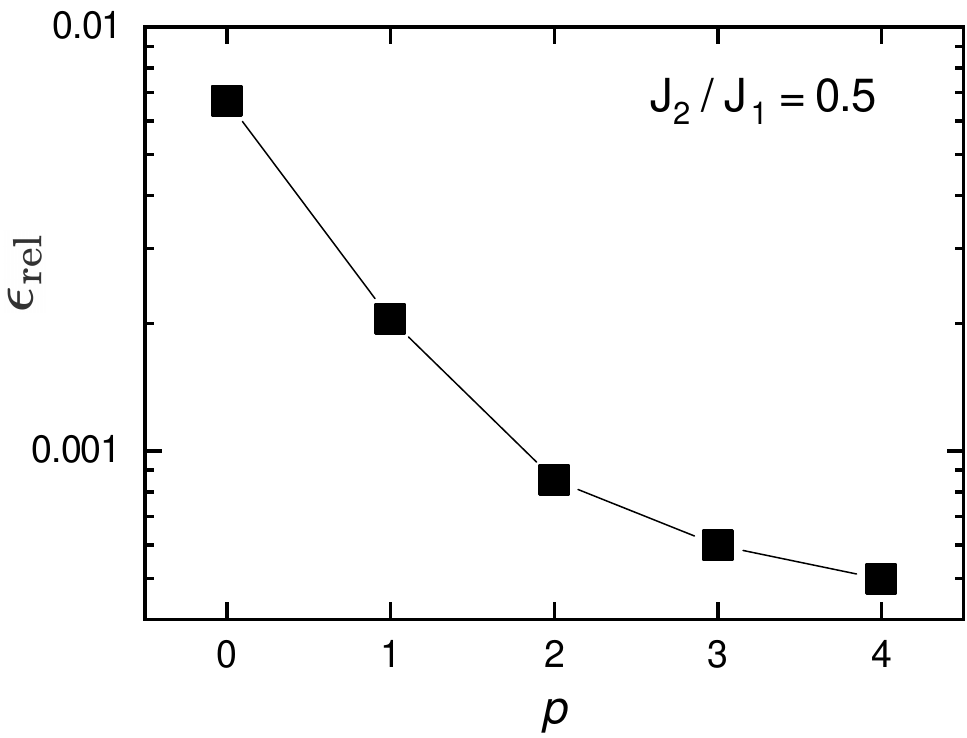}
  \caption{Improved energies on the 1D $L = 56$ chain with $J_2/J_1 = 0.5$. The vertical axis indicates the relative error of the improved energies with respect to the exact energy, $\epsilon_\text{rel} = (E - E_\text{exact}) / | E_\text{exact}|$.}
  \label{fig:1d_chain}
\end{figure}

To convincingly demonstrate the utility of the SLL algorithm, we apply the SLL algorithm to the Heisenberg  $J_1$-$J_2$ model on a 1D chain with $L = 56$, which has an exact solution $E_{\text{exact}} = -0.375 J_1$ for $J_2/J_1 = 0.5$. The results are shown in Figure~\ref{fig:1d_chain}. It can be observed that as $p$ increases, the relative error of the energy gradually decreases. Compared to $p = 2$ ($\epsilon_\text{rel} = 8.54 \times 10^{-4}$), the error for $p = 4$ ($\epsilon_\text{rel} = 4.98\times 10^{-4}$) decreased by approximately half, indicating that the computations for $p = 3$ and $4$ are effective. This suggests that as $p$ increases, the error continues to decrease.

\begin{table}[b!]
  \begin{center}    
    \caption{Sign accuracy of the NQS Lanczos method benchmarked against the exact solution on the $L=6$ square lattice, where the superposition state (with $p=2$) comprises three NQSs.}
    \label{tab:sign_accuracy}
    \setlength{\tabcolsep}{2.5mm}{
      \begin{tabular}{l r@{.}l r@{.}l }
        \hline
        \hline
        \multicolumn{1}{l}{State} & \multicolumn{2}{c}{$J_2/J_1 = 0.5$} & \multicolumn{2}{c}{$J_2/J_1 = 0.55$} \\
        \hline
        initial (MSR)                    & 0&9804       & 0&9352          \\
        superposition (SLL)        & 0&9983(1)    & 0&9960(1)          \\
        superposition (VMCL)    & 0&9981(1)    & 0&9955(1)            \\
        \hline
        \hline
      \end{tabular}
    }
  \end{center}
\end{table}

\begin{table*}[t!]
  \begin{center}
    \caption{Energies obtained from the NQS Lanczos method. SLL ($p = n$) represents the energy obtained by the SLL algorithm. VMCL ($p = n$, $\text{ANet}_j$) denotes the energies obtained by the optimization of $\text{ANet}_j$, which is the amplitude network in the $j$-th NQS constituting the superposition state. The result of MinSR is from Ref.~\cite{Chen2024MinSR}.}
    \label{tab:improved_energy}
    \setlength{\tabcolsep}{2.4mm}{
      \begin{tabular}{l l r@{.}l r@{.}l r@{.}l}
        \hline
        \hline
        \multicolumn{1}{l}{} & \multicolumn{1}{l}{Methods} & \multicolumn{2}{c}{$J_2/J_1=0.5$} & \multicolumn{2}{c}{$J_2/J_1=0.55$} & \multicolumn{2}{c}{$J_2/J_1=0.6$} \\
        \hline
        $L=6$  & SLL ($p=0$)             & -0&502184(2)  & -0&490953(3)  & -0&482273(4)        \\
               & SLL ($p=1$)             & -0&502918     & -0&492535     & -0&486399           \\
               & SLL ($p=2$)             & -0&503049     & -0&492820     & -0&487125           \\
               & VMCL ($p=1$, $\text{ANet}_1$) & -0&503052(3)  & -0&493142(5)  & -0&488024(7)        \\
               & VMCL ($p=2$, $\text{ANet}_1$) & -0&503119(4)  & -0&493357(5)  & -0&488569(7)        \\
               & VMCL ($p=2$, $\text{ANet}_1$ \& $\text{ANet}_2)$ & -0&503178(4)  & -0&493471(5)  & -0&488799(7)  \\
               & ED                      & -0&503810     & -0&495178     & -0&493239            \\
        \hline
        $L=10$ & SLL ($p=0$)             & -0&495627(6)  & -0&483490(5)  &  \multicolumn{2}{c}{} \\
               & SLL ($p=1$)             & -0&496064     & -0&484010     &  \multicolumn{2}{c}{} \\
               & VMCL ($p=1$, $\text{ANet}_1$) & -0&496102(7)  & -0&484811(8)  &  \multicolumn{2}{c}{} \\
               & MinSR                   & -0&497715(9)  &  \multicolumn{2}{c}{} &  \multicolumn{2}{c}{} \\
        \hline
        \hline
      \end{tabular}
    }
  \end{center}
\end{table*}

\subsection{Test of the VMC optimization part on the $6 \times 6$ and $10 \times 10$ lattices}
In the above tests, although the effectiveness of the combination of supervised learning with the Lanczos method has been demonstrated, the difficulty of loss convergence makes it impossible for the Lanczos state to be accurately represented by the NQS. In this section, we test the VMC optimization part on the superposition states obtained from the previous tests. For a superposition state $|\Psi\rangle$ that will be optimized by the VMC, all parameters of the sign networks and the superposition coefficients are fixed. The amplitude network $\text{ANet}_j$ ($j\neq0$) will be optimized one by one until the energy no longer decreases.

\begin{table}[b!]
  \begin{center}
    \caption{Coefficients $a_i$ and $b^2_i$ in the test of the NQS Lanczos method on the square lattices with $L=6$ and $L=10$.}
    \label{tab:a_b}
    \setlength{\tabcolsep}{1.2mm}{
      \begin{tabular}{l c r@{.}l r@{.}l r@{.}l}
        \hline
        \hline
        \multicolumn{1}{l}{} & \multicolumn{1}{l}{$J_2/J_1$} & \multicolumn{2}{c}{$0.5$} & \multicolumn{2}{c}{$0.55$} & \multicolumn{2}{c}{$0.6$} \\
        \hline
        $L=6$  & $a_0$   & -0&502184(2)   & -0&490953(3)   & -0&482273(4)          \\
               & $b^2_1$ & 0&000155(1)   & 0&000319(2)   & 0&000609(4)          \\
               & $a_1$   & -0&3253(3)     & -0&3407(3)     & -0&3722(2)            \\
               & $b^2_2$ & 0&0206(1)     & 0&0174(1)     & 0&0120(1)            \\
        \hline
        $L=10$ & $a_0$   & -0&495627(6)   & -0&483490(5)   &  \multicolumn{2}{c}{} \\
               & $b^2_1$ & 0&0000817(4)  & 0&0001135(6)  &  \multicolumn{2}{c}{} \\
        \hline
        \hline
      \end{tabular}
    }
  \end{center}
\end{table}

The results obtained by the VMC for the $L = 6$ and $L = 10$ square lattices are shown in Figs.~\ref{fig:L6} and~\ref{fig:L10}, respectively. In these figures, VMCL ($p = n$, $\text{ANet}_j$) denotes the energies obtained by the optimization of the superposition state, which consists of $n+1$ NQSs. $\text{ANet}_j$ indicate the $j$-th amplitude network is optimized. The specific energies are shown in Table~\ref{tab:improved_energy} of Appendix~\ref{sec:datas}. The energy improvement after the VMC is substantial. For $L = 6$, the improved energies achieved by optimizing only $\text{ANet}_1$ with $p = 1$ are better than those from SLL ($p = 2$). Moreover, the energy improvement increases further by optimizing $\text{ANet}_1$ and $\text{ANet}_2$ with $p = 2$.

The above results show that the VMC optimization part is very effective for the adjustment of the amplitude. We speculate that this adjustment partially compensates for the insufficient amplitude optimization during the supervised learning and further improves the generalization capability of the network.

We further benchmarked the sign accuracy of the superposition states against the exact solution for the $L = 6$ square lattice. The results are summarized in Table~\ref{tab:sign_accuracy}. The SLL algorithm significantly improves the sign accuracy from 98.04\% ($J_2/J_1=0.5$) and 93.52\% ($J_2/J_1=0.55$) of the initial states (with MSR imposed) to 99.83\% and 99.60\%, respectively, demonstrating its efficacy in refining the sign structure of initial states.

For the NQS Lanczos method, it might be sufficient to retain only those steps where significant energy improvement is observed, such as $p = 1$ or $p = 2$. Combining the SLL algorithm with the VMC optimization part, and using the optimized superposition state as the updated $|\psi^\mathrm{net}_0\rangle$, as shown in Fig.~\ref{fig:SLL_algorithm}, is a promising path to further improve the results. Due to limited GPU computational resources, we have not done it yet. We will try it in the future.

\section{Summary and outlook}
\label{sec:summary}
In summary, we integrate supervised learning, VMC, NQS, and the Lanczos method to develop a method, namely the NQS Lanczos method, to iteratively find the ground state of quantum many-body systems. It consists of two main parts: the supervised-learning Lanczos (SLL) algorithm and the VMC optimization part. This method is tested in high-frustration regions of the two-dimensional Heisenberg $J_1$-$J_2$ model on the square lattices with $L=4,6$, and $10$. The results demonstrate its effectiveness.

The main advantage of the NQS Lanczos method is that the computational cost increases linearly with the number of Lanczos steps. The SLL algorithm avoids the calculation of the expectation $\langle H^{2i+1} \rangle$ which was required in Ref.~\cite{chen2022lanczos}. This allows for more Lanczos steps to be performed with limited computational resources.

Currently, in this method, the accuracy of the supervised learning limits the efficiency of convergence. This limitation may be addressed through the following approaches: (i) increasing the size of the neural networks to enhance their expressive power; (ii) exploring new neural-network architectures that are better suited for two-dimensional lattice systems; (iii) designing improved loss functions to achieve better optimization.

\begin{acknowledgments}
This work was supported by the National Key R\&D Program of China (Grants No. 2024YFA1408602 and No. 2024YFA1408601) and the National Natural Science Foundation of China (Grant No. 12434009). Z.-Y.L. was also supported by the Innovation Program for Quantum Science and Technology (Grant No. 2021ZD0302402). Computational resources were provided by the Physical Laboratory of High Performance Computing in Renmin University of China.
\end{acknowledgments}

\section*{Code availability}
The code supporting the findings of this study is openly available in a GitHub repository via the URL provided in Ref.~\footnote{The code supporting the findings of this study is openly available in a GitHub repository at https://github.com/QTMEC-RUC/nqs-lanczos (will be publicly accessible upon publication of the paper).}.

\appendix

\section{Diagonalization of the Hamiltonian with a set of non-orthogonal basis states}
\label{sec:diagonalization}
For larger lattice sizes, the coefficients $a_i$ and $b_{i+1}$ are calculated approximately, and there are deviations between the NQS optimized by the supervised learning and the target state. The NQSs are not strictly orthogonal, and a Hamiltonian matrix constructed with these basis NQSs is not strictly tridiagonal. Therefore, it is necessary to transform the NQSs into an orthonormal basis set for diagonalization. The specific procedure is as follows.

For a set $\{ |\psi^\mathrm{net}_i\rangle \},i=0, \dots,p$, of NQSs obtained by the SLL algorithm, one can construct a Hermitian matrix $M$ whose element is
\begin{align}
    M_{ij} = \frac{\langle \psi^\mathrm{net}_i  | \psi^\mathrm{net}_j \rangle} {\langle \psi^\mathrm{net}_0 | \psi^\mathrm{net}_0 \rangle}.
\end{align}
Diagonalizing $M$ gives $ \Lambda = U^\dagger M U$ with eigenvalues $\lambda_i = \Lambda_{ii} > 0$. Define $S = U\sqrt{\Lambda^{-1}}$. An orthonormal basis set $\{ |\alpha_i \rangle\}$ can be obtained,
\begin{align}
    \{|\alpha_0\rangle, \cdots, |\alpha_{p}\rangle \} & = \{ |\psi^\mathrm{net}_0\rangle, \cdots, |\psi^\mathrm{net}_{p}\rangle \} S.
\end{align}
Define
\begin{align}
    H^{\psi}_{i, j} = \frac{\langle \psi^\mathrm{net}_i | H | \psi^\mathrm{net}_j \rangle} {\langle \psi^\mathrm{net}_0 | \psi^\mathrm{net}_0 \rangle}
\end{align}
and a Hermitian matrix $H^{\alpha} = S^\dagger H^\psi S$. Diagonalizing $H^{\alpha}$, one gets $H^{\alpha} \phi^\alpha = E \phi^\alpha$, 
where $E$ is the lowest eigenvalue, i.e., the improved energy, and $\phi^\alpha$ is the corresponding eigenvector and normalized to 1. Define $c = S \phi^\alpha$, then the superposition state
\begin{align}
    |\Psi\rangle = \sum_{i=0}^{p} c_i |\psi^\mathrm{net}_i\rangle 
    \label{eq:lc_NQSs}
\end{align}
is obtained, which is normalized to $\langle \psi^\mathrm{net}_0 | \psi^\mathrm{net}_0 \rangle$. The Hamiltonian expectation on $|\Psi\rangle$ is the improved energy $E$.

\section{Specific values in the test of the NQS Lanczos method}
\label{sec:datas}
This section presents specific values in the test of the NQS Lanczos method on square lattices with $L=6$ and $10$. The energies obtained by the SLL algorithm and the VMC optimization part are listed in Table~\ref{tab:improved_energy}. The coefficients $a_i$ and $b^2_i$ of the SLL algorithm are listed in Table~\ref{tab:a_b}.

\bibliography{reference}

\begin{thebibliography}{53}%
\makeatletter
\providecommand \@ifxundefined [1]{%
 \@ifx{#1\undefined}
}%
\providecommand \@ifnum [1]{%
 \ifnum #1\expandafter \@firstoftwo
 \else \expandafter \@secondoftwo
 \fi
}%
\providecommand \@ifx [1]{%
 \ifx #1\expandafter \@firstoftwo
 \else \expandafter \@secondoftwo
 \fi
}%
\providecommand \natexlab [1]{#1}%
\providecommand \enquote  [1]{``#1''}%
\providecommand \bibnamefont  [1]{#1}%
\providecommand \bibfnamefont [1]{#1}%
\providecommand \citenamefont [1]{#1}%
\providecommand \href@noop [0]{\@secondoftwo}%
\providecommand \href [0]{\begingroup \@sanitize@url \@href}%
\providecommand \@href[1]{\@@startlink{#1}\@@href}%
\providecommand \@@href[1]{\endgroup#1\@@endlink}%
\providecommand \@sanitize@url [0]{\catcode `\\12\catcode `\$12\catcode
  `\&12\catcode `\#12\catcode `\^12\catcode `\_12\catcode `\%12\relax}%
\providecommand \@@startlink[1]{}%
\providecommand \@@endlink[0]{}%
\providecommand \url  [0]{\begingroup\@sanitize@url \@url }%
\providecommand \@url [1]{\endgroup\@href {#1}{\urlprefix }}%
\providecommand \urlprefix  [0]{URL }%
\providecommand \Eprint [0]{\href }%
\providecommand \doibase [0]{https://doi.org/}%
\providecommand \selectlanguage [0]{\@gobble}%
\providecommand \bibinfo  [0]{\@secondoftwo}%
\providecommand \bibfield  [0]{\@secondoftwo}%
\providecommand \translation [1]{[#1]}%
\providecommand \BibitemOpen [0]{}%
\providecommand \bibitemStop [0]{}%
\providecommand \bibitemNoStop [0]{.\EOS\space}%
\providecommand \EOS [0]{\spacefactor3000\relax}%
\providecommand \BibitemShut  [1]{\csname bibitem#1\endcsname}%
\let\auto@bib@innerbib\@empty
\bibitem [{\citenamefont {Lou}\ and\ \citenamefont
  {Sandvik}(2007)}]{Sandvik2007VBBasis}%
  \BibitemOpen
  \bibfield  {author} {\bibinfo {author} {\bibfnamefont {J.}~\bibnamefont
  {Lou}}\ and\ \bibinfo {author} {\bibfnamefont {A.~W.}\ \bibnamefont
  {Sandvik}},\ }\bibfield  {title} {\bibinfo {title} {Variational ground states
  of two-dimensional antiferromagnets in the valence bond basis},\ }\href
  {https://doi.org/10.1103/PhysRevB.76.104432} {\bibfield  {journal} {\bibinfo
  {journal} {Phys. Rev. B}\ }\textbf {\bibinfo {volume} {76}},\ \bibinfo
  {pages} {104432} (\bibinfo {year} {2007})}\BibitemShut {NoStop}%
\bibitem [{\citenamefont {Sandvik}\ and\ \citenamefont
  {Evertz}(2010)}]{Sandvik2010_VBBasis+LoopUpdate}%
  \BibitemOpen
  \bibfield  {author} {\bibinfo {author} {\bibfnamefont {A.~W.}\ \bibnamefont
  {Sandvik}}\ and\ \bibinfo {author} {\bibfnamefont {H.~G.}\ \bibnamefont
  {Evertz}},\ }\bibfield  {title} {\bibinfo {title} {Loop updates for
  variational and projector quantum monte carlo simulations in the valence-bond
  basis},\ }\href {https://doi.org/10.1103/PhysRevB.82.024407} {\bibfield
  {journal} {\bibinfo  {journal} {Phys. Rev. B}\ }\textbf {\bibinfo {volume}
  {82}},\ \bibinfo {pages} {024407} (\bibinfo {year} {2010})}\BibitemShut
  {NoStop}%
\bibitem [{\citenamefont {Hu}\ \emph {et~al.}(2013)\citenamefont {Hu},
  \citenamefont {Becca}, \citenamefont {Parola},\ and\ \citenamefont
  {Sorella}}]{Sorella2013_VMC}%
  \BibitemOpen
  \bibfield  {author} {\bibinfo {author} {\bibfnamefont {W.-J.}\ \bibnamefont
  {Hu}}, \bibinfo {author} {\bibfnamefont {F.}~\bibnamefont {Becca}}, \bibinfo
  {author} {\bibfnamefont {A.}~\bibnamefont {Parola}},\ and\ \bibinfo {author}
  {\bibfnamefont {S.}~\bibnamefont {Sorella}},\ }\bibfield  {title} {\bibinfo
  {title} {Direct evidence for a gapless ${Z}_{2}$ spin liquid by frustrating
  {N\'eel} antiferromagnetism},\ }\href
  {https://doi.org/10.1103/PhysRevB.88.060402} {\bibfield  {journal} {\bibinfo
  {journal} {Phys. Rev. B}\ }\textbf {\bibinfo {volume} {88}},\ \bibinfo
  {pages} {060402(R)} (\bibinfo {year} {2013})}\BibitemShut {NoStop}%
\bibitem [{\citenamefont {Sandvik}\ and\ \citenamefont
  {Vidal}(2007)}]{Sandvik2007_VMC+TNS}%
  \BibitemOpen
  \bibfield  {author} {\bibinfo {author} {\bibfnamefont {A.~W.}\ \bibnamefont
  {Sandvik}}\ and\ \bibinfo {author} {\bibfnamefont {G.}~\bibnamefont
  {Vidal}},\ }\bibfield  {title} {\bibinfo {title} {Variational quantum monte
  carlo simulations with tensor-network states},\ }\href
  {https://doi.org/10.1103/PhysRevLett.99.220602} {\bibfield  {journal}
  {\bibinfo  {journal} {Phys. Rev. Lett.}\ }\textbf {\bibinfo {volume} {99}},\
  \bibinfo {pages} {220602} (\bibinfo {year} {2007})}\BibitemShut {NoStop}%
\bibitem [{\citenamefont {Tagliacozzo}\ \emph {et~al.}(2009)\citenamefont
  {Tagliacozzo}, \citenamefont {Evenbly},\ and\ \citenamefont
  {Vidal}}]{Tagliacozzo2009_TreeTensorNetwork}%
  \BibitemOpen
  \bibfield  {author} {\bibinfo {author} {\bibfnamefont {L.}~\bibnamefont
  {Tagliacozzo}}, \bibinfo {author} {\bibfnamefont {G.}~\bibnamefont
  {Evenbly}},\ and\ \bibinfo {author} {\bibfnamefont {G.}~\bibnamefont
  {Vidal}},\ }\bibfield  {title} {\bibinfo {title} {Simulation of
  two-dimensional quantum systems using a tree tensor network that exploits the
  entropic area law},\ }\href {https://doi.org/10.1103/PhysRevB.80.235127}
  {\bibfield  {journal} {\bibinfo  {journal} {Phys. Rev. B}\ }\textbf {\bibinfo
  {volume} {80}},\ \bibinfo {pages} {235127} (\bibinfo {year}
  {2009})}\BibitemShut {NoStop}%
\bibitem [{\citenamefont {Gong}\ \emph {et~al.}(2014)\citenamefont {Gong},
  \citenamefont {Zhu}, \citenamefont {Sheng}, \citenamefont {Motrunich},\ and\
  \citenamefont {Fisher}}]{GongShouShu2014_DMRG}%
  \BibitemOpen
  \bibfield  {author} {\bibinfo {author} {\bibfnamefont {S.-S.}\ \bibnamefont
  {Gong}}, \bibinfo {author} {\bibfnamefont {W.}~\bibnamefont {Zhu}}, \bibinfo
  {author} {\bibfnamefont {D.~N.}\ \bibnamefont {Sheng}}, \bibinfo {author}
  {\bibfnamefont {O.~I.}\ \bibnamefont {Motrunich}},\ and\ \bibinfo {author}
  {\bibfnamefont {M.~P.~A.}\ \bibnamefont {Fisher}},\ }\bibfield  {title}
  {\bibinfo {title} {Plaquette ordered phase and quantum phase diagram in the
  spin-$1/2$ ${J}_{1}\text{\ensuremath{-}}{J}_{2}$ square {Heisenberg} model},\
  }\href {https://doi.org/10.1103/PhysRevLett.113.027201} {\bibfield  {journal}
  {\bibinfo  {journal} {Phys. Rev. Lett.}\ }\textbf {\bibinfo {volume} {113}},\
  \bibinfo {pages} {027201} (\bibinfo {year} {2014})}\BibitemShut {NoStop}%
\bibitem [{\citenamefont {Stoudenmire}\ and\ \citenamefont
  {White}(2012)}]{Stoudenmire2012_2DsystemDMRG}%
  \BibitemOpen
  \bibfield  {author} {\bibinfo {author} {\bibfnamefont {E.}~\bibnamefont
  {Stoudenmire}}\ and\ \bibinfo {author} {\bibfnamefont {S.~R.}\ \bibnamefont
  {White}},\ }\bibfield  {title} {\bibinfo {title} {Studying two-dimensional
  systems with the density matrix renormalization group},\ }\href
  {https://doi.org/10.1146/annurev-conmatphys-020911-125018} {\bibfield
  {journal} {\bibinfo  {journal} {Annual Review of Condensed Matter Physics}\
  }\textbf {\bibinfo {volume} {3}},\ \bibinfo {pages} {111} (\bibinfo {year}
  {2012})}\BibitemShut {NoStop}%
\bibitem [{\citenamefont {Wang}\ and\ \citenamefont
  {Sandvik}(2018)}]{Sandvik2018_DMRG}%
  \BibitemOpen
  \bibfield  {author} {\bibinfo {author} {\bibfnamefont {L.}~\bibnamefont
  {Wang}}\ and\ \bibinfo {author} {\bibfnamefont {A.~W.}\ \bibnamefont
  {Sandvik}},\ }\bibfield  {title} {\bibinfo {title} {Critical level crossings
  and gapless spin liquid in the square-lattice spin-$1/2$
  ${J}_{1}\ensuremath{-}{J}_{2}$ {Heisenberg} antiferromagnet},\ }\href
  {https://doi.org/10.1103/PhysRevLett.121.107202} {\bibfield  {journal}
  {\bibinfo  {journal} {Phys. Rev. Lett.}\ }\textbf {\bibinfo {volume} {121}},\
  \bibinfo {pages} {107202} (\bibinfo {year} {2018})}\BibitemShut {NoStop}%
\bibitem [{\citenamefont {Sandvik}(1997)}]{Sandvik1997SSEQMC}%
  \BibitemOpen
  \bibfield  {author} {\bibinfo {author} {\bibfnamefont {A.~W.}\ \bibnamefont
  {Sandvik}},\ }\bibfield  {title} {\bibinfo {title} {Finite-size scaling of
  the ground-state parameters of the two-dimensional {Heisenberg} model},\
  }\href {https://doi.org/10.1103/PhysRevB.56.11678} {\bibfield  {journal}
  {\bibinfo  {journal} {Phys. Rev. B}\ }\textbf {\bibinfo {volume} {56}},\
  \bibinfo {pages} {11678} (\bibinfo {year} {1997})}\BibitemShut {NoStop}%
\bibitem [{\citenamefont {Carrasquilla}\ and\ \citenamefont
  {Torlai}(2021)}]{Juan2021PRX_NQSReview}%
  \BibitemOpen
  \bibfield  {author} {\bibinfo {author} {\bibfnamefont {J.}~\bibnamefont
  {Carrasquilla}}\ and\ \bibinfo {author} {\bibfnamefont {G.}~\bibnamefont
  {Torlai}},\ }\bibfield  {title} {\bibinfo {title} {How to use neural networks
  to investigate quantum many-body physics},\ }\href
  {https://doi.org/10.1103/PRXQuantum.2.040201} {\bibfield  {journal} {\bibinfo
   {journal} {PRX Quantum}\ }\textbf {\bibinfo {volume} {2}},\ \bibinfo {pages}
  {040201} (\bibinfo {year} {2021})}\BibitemShut {NoStop}%
\bibitem [{\citenamefont {Nomura}(2023)}]{Nomura2023RBMNQSReview}%
  \BibitemOpen
  \bibfield  {author} {\bibinfo {author} {\bibfnamefont {Y.}~\bibnamefont
  {Nomura}},\ }\bibfield  {title} {\bibinfo {title} {Boltzmann machines and
  quantum many-body problems},\ }\href
  {https://api.semanticscholar.org/CorpusID:259286933} {\bibfield  {journal}
  {\bibinfo  {journal} {Journal of Physics: Condensed Matter}\ }\textbf
  {\bibinfo {volume} {36}} (\bibinfo {year} {2023})}\BibitemShut {NoStop}%
\bibitem [{\citenamefont {Hermann}\ \emph {et~al.}(2023)\citenamefont
  {Hermann}, \citenamefont {Spencer}, \citenamefont {Choo}, \citenamefont
  {Mezzacapo}, \citenamefont {Foulkes}, \citenamefont {Pfau}, \citenamefont
  {Carleo},\ and\ \citenamefont {Noé}}]{Hermann2023_NQSChemistryReview}%
  \BibitemOpen
  \bibfield  {author} {\bibinfo {author} {\bibfnamefont {J.}~\bibnamefont
  {Hermann}}, \bibinfo {author} {\bibfnamefont {J.}~\bibnamefont {Spencer}},
  \bibinfo {author} {\bibfnamefont {K.}~\bibnamefont {Choo}}, \bibinfo {author}
  {\bibfnamefont {A.}~\bibnamefont {Mezzacapo}}, \bibinfo {author}
  {\bibfnamefont {W.~M.~C.}\ \bibnamefont {Foulkes}}, \bibinfo {author}
  {\bibfnamefont {D.}~\bibnamefont {Pfau}}, \bibinfo {author} {\bibfnamefont
  {G.}~\bibnamefont {Carleo}},\ and\ \bibinfo {author} {\bibfnamefont
  {F.}~\bibnamefont {Noé}},\ }\bibfield  {title} {\bibinfo {title} {Ab initio
  quantum chemistry with neural-network wavefunctions},\ }\href
  {https://doi.org/10.1038/s41570-023-00516-8} {\bibfield  {journal} {\bibinfo
  {journal} {Nature Reviews Chemistry}\ }\textbf {\bibinfo {volume} {7}},\
  \bibinfo {pages} {692} (\bibinfo {year} {2023})}\BibitemShut {NoStop}%
\bibitem [{\citenamefont {Medvidović}\ and\ \citenamefont
  {Moreno}(2024)}]{Medvidovic2024NQSReview}%
  \BibitemOpen
  \bibfield  {author} {\bibinfo {author} {\bibfnamefont {M.}~\bibnamefont
  {Medvidović}}\ and\ \bibinfo {author} {\bibfnamefont {J.~R.}\ \bibnamefont
  {Moreno}},\ }\bibfield  {title} {\bibinfo {title} {Neural-network quantum
  states for many-body physics},\ }\href
  {https://doi.org/10.1140/epjp/s13360-024-05311-y} {\bibfield  {journal}
  {\bibinfo  {journal} {The European Physical Journal Plus}\ }\textbf {\bibinfo
  {volume} {139}},\ \bibinfo {pages} {631} (\bibinfo {year}
  {2024})}\BibitemShut {NoStop}%
\bibitem [{\citenamefont {Lange}\ \emph
  {et~al.}(2024{\natexlab{a}})\citenamefont {Lange}, \citenamefont {Van~de
  Walle}, \citenamefont {Abedinnia},\ and\ \citenamefont
  {Bohrdt}}]{Lange2024NQSReview}%
  \BibitemOpen
  \bibfield  {author} {\bibinfo {author} {\bibfnamefont {H.}~\bibnamefont
  {Lange}}, \bibinfo {author} {\bibfnamefont {A.}~\bibnamefont {Van~de Walle}},
  \bibinfo {author} {\bibfnamefont {A.}~\bibnamefont {Abedinnia}},\ and\
  \bibinfo {author} {\bibfnamefont {A.}~\bibnamefont {Bohrdt}},\ }\bibfield
  {title} {\bibinfo {title} {From architectures to applications: a review of
  neural quantum states},\ }\href {https://doi.org/10.1088/2058-9565/ad7168}
  {\bibfield  {journal} {\bibinfo  {journal} {Quantum Science and Technology}\
  }\textbf {\bibinfo {volume} {9}},\ \bibinfo {pages} {040501} (\bibinfo {year}
  {2024}{\natexlab{a}})}\BibitemShut {NoStop}%
\bibitem [{\citenamefont {Carleo}\ and\ \citenamefont
  {Troyer}(2017)}]{GiuseppeCarleo2017_RBM}%
  \BibitemOpen
  \bibfield  {author} {\bibinfo {author} {\bibfnamefont {G.}~\bibnamefont
  {Carleo}}\ and\ \bibinfo {author} {\bibfnamefont {M.}~\bibnamefont
  {Troyer}},\ }\bibfield  {title} {\bibinfo {title} {Solving the quantum
  many-body problem with artificial neural networks},\ }\href
  {https://doi.org/10.1126/science.aag2302} {\bibfield  {journal} {\bibinfo
  {journal} {Science}\ }\textbf {\bibinfo {volume} {355}},\ \bibinfo {pages}
  {602} (\bibinfo {year} {2017})}\BibitemShut {NoStop}%
\bibitem [{\citenamefont {Sorella}\ \emph {et~al.}(2007)\citenamefont
  {Sorella}, \citenamefont {Casula},\ and\ \citenamefont
  {Rocca}}]{Sorella2007SR}%
  \BibitemOpen
  \bibfield  {author} {\bibinfo {author} {\bibfnamefont {S.}~\bibnamefont
  {Sorella}}, \bibinfo {author} {\bibfnamefont {M.}~\bibnamefont {Casula}},\
  and\ \bibinfo {author} {\bibfnamefont {D.}~\bibnamefont {Rocca}},\ }\bibfield
   {title} {\bibinfo {title} {Weak binding between two aromatic rings: Feeling
  the van der {Waals} attraction by quantum monte carlo methods},\ }\bibfield
  {journal} {\bibinfo  {journal} {J. Chem. Phys}\ }\textbf {\bibinfo {volume}
  {127}},\ \href {https://doi.org/10.1063/1.2746035} {10.1063/1.2746035}
  (\bibinfo {year} {2007})\BibitemShut {NoStop}%
\bibitem [{\citenamefont {Szab\'o}\ and\ \citenamefont
  {Castelnovo}(2020)}]{Claudio2020_NQS_SignProblem}%
  \BibitemOpen
  \bibfield  {author} {\bibinfo {author} {\bibfnamefont {A.}~\bibnamefont
  {Szab\'o}}\ and\ \bibinfo {author} {\bibfnamefont {C.}~\bibnamefont
  {Castelnovo}},\ }\bibfield  {title} {\bibinfo {title} {Neural network wave
  functions and the sign problem},\ }\href
  {https://doi.org/10.1103/PhysRevResearch.2.033075} {\bibfield  {journal}
  {\bibinfo  {journal} {Phys. Rev. Res.}\ }\textbf {\bibinfo {volume} {2}},\
  \bibinfo {pages} {033075} (\bibinfo {year} {2020})}\BibitemShut {NoStop}%
\bibitem [{\citenamefont {Westerhout}\ \emph {et~al.}(2023)\citenamefont
  {Westerhout}, \citenamefont {Katsnelson},\ and\ \citenamefont
  {Bagrov}}]{westerhout2023SignStructure}%
  \BibitemOpen
  \bibfield  {author} {\bibinfo {author} {\bibfnamefont {T.}~\bibnamefont
  {Westerhout}}, \bibinfo {author} {\bibfnamefont {M.~I.}\ \bibnamefont
  {Katsnelson}},\ and\ \bibinfo {author} {\bibfnamefont {A.~A.}\ \bibnamefont
  {Bagrov}},\ }\bibfield  {title} {\bibinfo {title} {Many-body quantum sign
  structures as non-glassy ising models},\ }\href
  {https://doi.org/10.1038/s42005-023-01388-6} {\bibfield  {journal} {\bibinfo
  {journal} {Commun. Phys.}\ }\textbf {\bibinfo {volume} {6}},\ \bibinfo
  {pages} {275} (\bibinfo {year} {2023})}\BibitemShut {NoStop}%
\bibitem [{\citenamefont {Westerhout}\ \emph {et~al.}(2020)\citenamefont
  {Westerhout}, \citenamefont {Astrakhantsev}, \citenamefont {Tikhonov},
  \citenamefont {Katsnelson},\ and\ \citenamefont
  {Bagrov}}]{Westerhout2020_difference_Sign_amp}%
  \BibitemOpen
  \bibfield  {author} {\bibinfo {author} {\bibfnamefont {T.}~\bibnamefont
  {Westerhout}}, \bibinfo {author} {\bibfnamefont {N.}~\bibnamefont
  {Astrakhantsev}}, \bibinfo {author} {\bibfnamefont {K.~S.}\ \bibnamefont
  {Tikhonov}}, \bibinfo {author} {\bibfnamefont {M.~I.}\ \bibnamefont
  {Katsnelson}},\ and\ \bibinfo {author} {\bibfnamefont {A.~A.}\ \bibnamefont
  {Bagrov}},\ }\bibfield  {title} {\bibinfo {title} {Generalization properties
  of neural network approximations to frustrated magnet ground states},\ }\href
  {https://doi.org/10.1038/s41467-020-15402-w} {\bibfield  {journal} {\bibinfo
  {journal} {Nat. Commun.}\ }\textbf {\bibinfo {volume} {11}},\ \bibinfo
  {pages} {1593} (\bibinfo {year} {2020})}\BibitemShut {NoStop}%
\bibitem [{\citenamefont {Chen}\ \emph
  {et~al.}(2022{\natexlab{a}})\citenamefont {Chen}, \citenamefont {Choo},
  \citenamefont {Astrakhantsev},\ and\ \citenamefont
  {Neupert}}]{Chen2022_SignStructure-NeuralNetwork}%
  \BibitemOpen
  \bibfield  {author} {\bibinfo {author} {\bibfnamefont {A.}~\bibnamefont
  {Chen}}, \bibinfo {author} {\bibfnamefont {K.}~\bibnamefont {Choo}}, \bibinfo
  {author} {\bibfnamefont {N.}~\bibnamefont {Astrakhantsev}},\ and\ \bibinfo
  {author} {\bibfnamefont {T.}~\bibnamefont {Neupert}},\ }\bibfield  {title}
  {\bibinfo {title} {Neural network evolution strategy for solving quantum sign
  structures},\ }\href {https://doi.org/10.1103/PhysRevResearch.4.L022026}
  {\bibfield  {journal} {\bibinfo  {journal} {Phys. Rev. Res.}\ }\textbf
  {\bibinfo {volume} {4}},\ \bibinfo {pages} {L022026} (\bibinfo {year}
  {2022}{\natexlab{a}})}\BibitemShut {NoStop}%
\bibitem [{\citenamefont {Cai}\ and\ \citenamefont {Liu}(2018)}]{Cai2018_ANN}%
  \BibitemOpen
  \bibfield  {author} {\bibinfo {author} {\bibfnamefont {Z.}~\bibnamefont
  {Cai}}\ and\ \bibinfo {author} {\bibfnamefont {J.}~\bibnamefont {Liu}},\
  }\bibfield  {title} {\bibinfo {title} {Approximating quantum many-body wave
  functions using artificial neural networks},\ }\href
  {https://doi.org/10.1103/PhysRevB.97.035116} {\bibfield  {journal} {\bibinfo
  {journal} {Phys. Rev. B}\ }\textbf {\bibinfo {volume} {97}},\ \bibinfo
  {pages} {035116} (\bibinfo {year} {2018})}\BibitemShut {NoStop}%
\bibitem [{\citenamefont {Pfau}\ \emph {et~al.}(2020)\citenamefont {Pfau},
  \citenamefont {Spencer}, \citenamefont {Matthews},\ and\ \citenamefont
  {Foulkes}}]{Pfau2020Ferminet}%
  \BibitemOpen
  \bibfield  {author} {\bibinfo {author} {\bibfnamefont {D.}~\bibnamefont
  {Pfau}}, \bibinfo {author} {\bibfnamefont {J.~S.}\ \bibnamefont {Spencer}},
  \bibinfo {author} {\bibfnamefont {A.~G. D.~G.}\ \bibnamefont {Matthews}},\
  and\ \bibinfo {author} {\bibfnamefont {W.~M.~C.}\ \bibnamefont {Foulkes}},\
  }\bibfield  {title} {\bibinfo {title} {Ab initio solution of the
  many-electron {Schr\"odinger} equation with deep neural networks},\ }\href
  {https://doi.org/10.1103/PhysRevResearch.2.033429} {\bibfield  {journal}
  {\bibinfo  {journal} {Phys. Rev. Res.}\ }\textbf {\bibinfo {volume} {2}},\
  \bibinfo {pages} {033429} (\bibinfo {year} {2020})}\BibitemShut {NoStop}%
\bibitem [{\citenamefont {Hibat-Allah}\ \emph {et~al.}(2020)\citenamefont
  {Hibat-Allah}, \citenamefont {Ganahl}, \citenamefont {Hayward}, \citenamefont
  {Melko},\ and\ \citenamefont {Carrasquilla}}]{HibatAllah2020_RNN_NQS}%
  \BibitemOpen
  \bibfield  {author} {\bibinfo {author} {\bibfnamefont {M.}~\bibnamefont
  {Hibat-Allah}}, \bibinfo {author} {\bibfnamefont {M.}~\bibnamefont {Ganahl}},
  \bibinfo {author} {\bibfnamefont {L.~E.}\ \bibnamefont {Hayward}}, \bibinfo
  {author} {\bibfnamefont {R.~G.}\ \bibnamefont {Melko}},\ and\ \bibinfo
  {author} {\bibfnamefont {J.}~\bibnamefont {Carrasquilla}},\ }\bibfield
  {title} {\bibinfo {title} {Recurrent neural network wave functions},\ }\href
  {https://doi.org/10.1103/PhysRevResearch.2.023358} {\bibfield  {journal}
  {\bibinfo  {journal} {Phys. Rev. Res.}\ }\textbf {\bibinfo {volume} {2}},\
  \bibinfo {pages} {023358} (\bibinfo {year} {2020})}\BibitemShut {NoStop}%
\bibitem [{\citenamefont {Kochkov}\ \emph {et~al.}(2021)\citenamefont
  {Kochkov}, \citenamefont {Pfaff}, \citenamefont {Sanchez-Gonzalez},
  \citenamefont {Battaglia},\ and\ \citenamefont {Clark}}]{kochkov2021GNN}%
  \BibitemOpen
  \bibfield  {author} {\bibinfo {author} {\bibfnamefont {D.}~\bibnamefont
  {Kochkov}}, \bibinfo {author} {\bibfnamefont {T.}~\bibnamefont {Pfaff}},
  \bibinfo {author} {\bibfnamefont {A.}~\bibnamefont {Sanchez-Gonzalez}},
  \bibinfo {author} {\bibfnamefont {P.}~\bibnamefont {Battaglia}},\ and\
  \bibinfo {author} {\bibfnamefont {B.~K.}\ \bibnamefont {Clark}},\ }\bibfield
  {title} {\bibinfo {title} {Learning ground states of quantum {Hamiltonians}
  with graph networks},\ }\href {https://doi.org/10.48550/arXiv.2110.06390}
  {\bibfield  {journal} {\bibinfo  {journal} {arXiv:2110.06390}\ } (\bibinfo
  {year} {2021})}\BibitemShut {NoStop}%
\bibitem [{\citenamefont {Fu}\ \emph {et~al.}(2022)\citenamefont {Fu},
  \citenamefont {Zhang}, \citenamefont {Zhang}, \citenamefont {Ling},
  \citenamefont {Xu},\ and\ \citenamefont {Ji}}]{fu2022latticeCNN}%
  \BibitemOpen
  \bibfield  {author} {\bibinfo {author} {\bibfnamefont {C.}~\bibnamefont
  {Fu}}, \bibinfo {author} {\bibfnamefont {X.}~\bibnamefont {Zhang}}, \bibinfo
  {author} {\bibfnamefont {H.}~\bibnamefont {Zhang}}, \bibinfo {author}
  {\bibfnamefont {H.}~\bibnamefont {Ling}}, \bibinfo {author} {\bibfnamefont
  {S.}~\bibnamefont {Xu}},\ and\ \bibinfo {author} {\bibfnamefont
  {S.}~\bibnamefont {Ji}},\ }\bibfield  {title} {\bibinfo {title} {Lattice
  convolutional networks for learning ground states of quantum many-body
  systems},\ }\href {https://doi.org/10.48550/arXiv.2206.07370} {\bibfield
  {journal} {\bibinfo  {journal} {arXiv:2206.07370}\ } (\bibinfo {year}
  {2022})}\BibitemShut {NoStop}%
\bibitem [{\citenamefont {Roth}\ \emph {et~al.}(2023)\citenamefont {Roth},
  \citenamefont {Szab\'o},\ and\ \citenamefont {MacDonald}}]{roth2023_GCNN}%
  \BibitemOpen
  \bibfield  {author} {\bibinfo {author} {\bibfnamefont {C.}~\bibnamefont
  {Roth}}, \bibinfo {author} {\bibfnamefont {A.}~\bibnamefont {Szab\'o}},\ and\
  \bibinfo {author} {\bibfnamefont {A.~H.}\ \bibnamefont {MacDonald}},\
  }\bibfield  {title} {\bibinfo {title} {High-accuracy variational monte carlo
  for frustrated magnets with deep neural networks},\ }\href
  {https://doi.org/10.1103/PhysRevB.108.054410} {\bibfield  {journal} {\bibinfo
   {journal} {Phys. Rev. B}\ }\textbf {\bibinfo {volume} {108}},\ \bibinfo
  {pages} {054410} (\bibinfo {year} {2023})}\BibitemShut {NoStop}%
\bibitem [{\citenamefont {Rende}\ \emph {et~al.}(2024)\citenamefont {Rende},
  \citenamefont {Viteritti}, \citenamefont {Bardone}, \citenamefont {Becca},\
  and\ \citenamefont {Goldt}}]{Rende2024DeepViT}%
  \BibitemOpen
  \bibfield  {author} {\bibinfo {author} {\bibfnamefont {R.}~\bibnamefont
  {Rende}}, \bibinfo {author} {\bibfnamefont {L.~L.}\ \bibnamefont
  {Viteritti}}, \bibinfo {author} {\bibfnamefont {L.}~\bibnamefont {Bardone}},
  \bibinfo {author} {\bibfnamefont {F.}~\bibnamefont {Becca}},\ and\ \bibinfo
  {author} {\bibfnamefont {S.}~\bibnamefont {Goldt}},\ }\bibfield  {title}
  {\bibinfo {title} {A simple linear algebra identity to optimize large-scale
  neural network quantum states},\ }\href
  {https://doi.org/10.1038/s42005-024-01732-4} {\bibfield  {journal} {\bibinfo
  {journal} {Communications Physics}\ }\textbf {\bibinfo {volume} {7}},\
  \bibinfo {pages} {260} (\bibinfo {year} {2024})}\BibitemShut {NoStop}%
\bibitem [{\citenamefont {Lange}\ \emph
  {et~al.}(2024{\natexlab{b}})\citenamefont {Lange}, \citenamefont {Döschl},
  \citenamefont {Carrasquilla},\ and\ \citenamefont
  {Bohrdt}}]{Lange2024RNN_NQS}%
  \BibitemOpen
  \bibfield  {author} {\bibinfo {author} {\bibfnamefont {H.}~\bibnamefont
  {Lange}}, \bibinfo {author} {\bibfnamefont {F.}~\bibnamefont {Döschl}},
  \bibinfo {author} {\bibfnamefont {J.}~\bibnamefont {Carrasquilla}},\ and\
  \bibinfo {author} {\bibfnamefont {A.}~\bibnamefont {Bohrdt}},\ }\bibfield
  {title} {\bibinfo {title} {Neural network approach to quasiparticle
  dispersions in doped antiferromagnets},\ }\href
  {https://doi.org/10.1038/s42005-024-01678-7} {\bibfield  {journal} {\bibinfo
  {journal} {Communications Physics}\ }\textbf {\bibinfo {volume} {7}},\
  \bibinfo {pages} {187} (\bibinfo {year} {2024}{\natexlab{b}})}\BibitemShut
  {NoStop}%
\bibitem [{\citenamefont {Chen}\ and\ \citenamefont
  {Heyl}(2024)}]{Chen2024MinSR}%
  \BibitemOpen
  \bibfield  {author} {\bibinfo {author} {\bibfnamefont {A.}~\bibnamefont
  {Chen}}\ and\ \bibinfo {author} {\bibfnamefont {M.}~\bibnamefont {Heyl}},\
  }\bibfield  {title} {\bibinfo {title} {Empowering deep neural quantum states
  through efficient optimization},\ }\href
  {https://doi.org/10.1038/s41567-024-02566-1} {\bibfield  {journal} {\bibinfo
  {journal} {Nature Physics}\ }\textbf {\bibinfo {volume} {20}},\ \bibinfo
  {pages} {1476} (\bibinfo {year} {2024})}\BibitemShut {NoStop}%
\bibitem [{\citenamefont {Drissi}\ \emph {et~al.}(2024)\citenamefont {Drissi},
  \citenamefont {Keeble}, \citenamefont {Rozalén~Sarmiento},\ and\
  \citenamefont {Rios}}]{Drissi2024SecondOrderOptimization}%
  \BibitemOpen
  \bibfield  {author} {\bibinfo {author} {\bibfnamefont {M.}~\bibnamefont
  {Drissi}}, \bibinfo {author} {\bibfnamefont {J.~W.~T.}\ \bibnamefont
  {Keeble}}, \bibinfo {author} {\bibfnamefont {J.}~\bibnamefont
  {Rozalén~Sarmiento}},\ and\ \bibinfo {author} {\bibfnamefont
  {A.}~\bibnamefont {Rios}},\ }\bibfield  {title} {\bibinfo {title}
  {Second-order optimization strategies for neural network quantum states},\
  }\href {https://doi.org/10.1098/rsta.2024.0057} {\bibfield  {journal}
  {\bibinfo  {journal} {Philosophical Transactions of the Royal Society A:
  Mathematical, Physical and Engineering Sciences}\ }\textbf {\bibinfo {volume}
  {382}},\ \bibinfo {pages} {20240057} (\bibinfo {year} {2024})}\BibitemShut
  {NoStop}%
\bibitem [{\citenamefont {Lanczos}(1952)}]{Lanczos1952Method}%
  \BibitemOpen
  \bibfield  {author} {\bibinfo {author} {\bibfnamefont {C.}~\bibnamefont
  {Lanczos}},\ }\bibfield  {title} {\bibinfo {title} {Solution of systems of
  linear equations by minimized iterations1},\ }\href
  {https://api.semanticscholar.org/CorpusID:7484650} {\bibfield  {journal}
  {\bibinfo  {journal} {Journal of research of the National Bureau of
  Standards}\ }\textbf {\bibinfo {volume} {49}},\ \bibinfo {pages} {33}
  (\bibinfo {year} {1952})}\BibitemShut {NoStop}%
\bibitem [{\citenamefont {Sorella}(2001)}]{sorella2001VMC_Lanczos}%
  \BibitemOpen
  \bibfield  {author} {\bibinfo {author} {\bibfnamefont {S.}~\bibnamefont
  {Sorella}},\ }\bibfield  {title} {\bibinfo {title} {Generalized {Lanczos}
  algorithm for variational quantum monte carlo},\ }\href
  {https://doi.org/10.1103/PhysRevB.64.024512} {\bibfield  {journal} {\bibinfo
  {journal} {Phys. Rev. B}\ }\textbf {\bibinfo {volume} {64}},\ \bibinfo
  {pages} {024512} (\bibinfo {year} {2001})}\BibitemShut {NoStop}%
\bibitem [{\citenamefont {Huang}\ \emph {et~al.}(2018)\citenamefont {Huang},
  \citenamefont {Liao}, \citenamefont {Liu}, \citenamefont {Xie}, \citenamefont
  {Xie}, \citenamefont {Zhao}, \citenamefont {Chen},\ and\ \citenamefont
  {Xiang}}]{huang2018tensornetwork_Lanczos}%
  \BibitemOpen
  \bibfield  {author} {\bibinfo {author} {\bibfnamefont {R.-Z.}\ \bibnamefont
  {Huang}}, \bibinfo {author} {\bibfnamefont {H.-J.}\ \bibnamefont {Liao}},
  \bibinfo {author} {\bibfnamefont {Z.-Y.}\ \bibnamefont {Liu}}, \bibinfo
  {author} {\bibfnamefont {H.-D.}\ \bibnamefont {Xie}}, \bibinfo {author}
  {\bibfnamefont {Z.-Y.}\ \bibnamefont {Xie}}, \bibinfo {author} {\bibfnamefont
  {H.-H.}\ \bibnamefont {Zhao}}, \bibinfo {author} {\bibfnamefont
  {J.}~\bibnamefont {Chen}},\ and\ \bibinfo {author} {\bibfnamefont
  {T.}~\bibnamefont {Xiang}},\ }\bibfield  {title} {\bibinfo {title}
  {Generalized {Lanczos} method for systematic optimization of tensor network
  states},\ }\href {https://doi.org/10.1088/1674-1056/27/7/070501} {\bibfield
  {journal} {\bibinfo  {journal} {Chinese Physics B}\ }\textbf {\bibinfo
  {volume} {27}},\ \bibinfo {pages} {070501} (\bibinfo {year}
  {2018})}\BibitemShut {NoStop}%
\bibitem [{\citenamefont {Chen}\ \emph
  {et~al.}(2022{\natexlab{b}})\citenamefont {Chen}, \citenamefont {Hendry},
  \citenamefont {Weinberg},\ and\ \citenamefont {Feiguin}}]{chen2022lanczos}%
  \BibitemOpen
  \bibfield  {author} {\bibinfo {author} {\bibfnamefont {H.}~\bibnamefont
  {Chen}}, \bibinfo {author} {\bibfnamefont {D.~G.}\ \bibnamefont {Hendry}},
  \bibinfo {author} {\bibfnamefont {P.~E.}\ \bibnamefont {Weinberg}},\ and\
  \bibinfo {author} {\bibfnamefont {A.}~\bibnamefont {Feiguin}},\ }\bibfield
  {title} {\bibinfo {title} {Systematic improvement of neural network quantum
  states using {Lanczos}},\ }in\ \href
  {https://openreview.net/forum?id=qZUHvvtbzy} {\emph {\bibinfo {booktitle}
  {Advances in Neural Information Processing Systems}}},\ \bibinfo {editor}
  {edited by\ \bibinfo {editor} {\bibfnamefont {A.~H.}\ \bibnamefont {Oh}},
  \bibinfo {editor} {\bibfnamefont {A.}~\bibnamefont {Agarwal}}, \bibinfo
  {editor} {\bibfnamefont {D.}~\bibnamefont {Belgrave}},\ and\ \bibinfo
  {editor} {\bibfnamefont {K.}~\bibnamefont {Cho}}}\ (\bibinfo {year}
  {2022})\BibitemShut {NoStop}%
\bibitem [{\citenamefont {Wang}\ \emph {et~al.}(2024)\citenamefont {Wang},
  \citenamefont {Wu}, \citenamefont {He},\ and\ \citenamefont
  {Lu}}]{Jqwang2024aCNN}%
  \BibitemOpen
  \bibfield  {author} {\bibinfo {author} {\bibfnamefont {J.-Q.}\ \bibnamefont
  {Wang}}, \bibinfo {author} {\bibfnamefont {H.-Q.}\ \bibnamefont {Wu}},
  \bibinfo {author} {\bibfnamefont {R.-Q.}\ \bibnamefont {He}},\ and\ \bibinfo
  {author} {\bibfnamefont {Z.-Y.}\ \bibnamefont {Lu}},\ }\bibfield  {title}
  {\bibinfo {title} {Variational optimization of the amplitude of
  neural-network quantum many-body ground states},\ }\href
  {https://doi.org/10.1103/PhysRevB.109.245120} {\bibfield  {journal} {\bibinfo
   {journal} {Phys. Rev. B}\ }\textbf {\bibinfo {volume} {109}},\ \bibinfo
  {pages} {245120} (\bibinfo {year} {2024})}\BibitemShut {NoStop}%
\bibitem [{\citenamefont {Salakhutdinov}\ and\ \citenamefont
  {Hinton}(2009)}]{Salakhutdinov2009DeepRBM}%
  \BibitemOpen
  \bibfield  {author} {\bibinfo {author} {\bibfnamefont {R.}~\bibnamefont
  {Salakhutdinov}}\ and\ \bibinfo {author} {\bibfnamefont {G.~E.}\ \bibnamefont
  {Hinton}},\ }\bibfield  {title} {\bibinfo {title} {Deep {Boltzmann}
  machines},\ }in\ \href {https://api.semanticscholar.org/CorpusID:877639}
  {\emph {\bibinfo {booktitle} {International Conference on Artificial
  Intelligence and Statistics}}}\ (\bibinfo {year} {2009})\BibitemShut
  {NoStop}%
\bibitem [{\citenamefont {Rumelhart}\ \emph {et~al.}(1986)\citenamefont
  {Rumelhart}, \citenamefont {Hinton},\ and\ \citenamefont
  {Williams}}]{Rumelhart1986MLP}%
  \BibitemOpen
  \bibfield  {author} {\bibinfo {author} {\bibfnamefont {D.~E.}\ \bibnamefont
  {Rumelhart}}, \bibinfo {author} {\bibfnamefont {G.~E.}\ \bibnamefont
  {Hinton}},\ and\ \bibinfo {author} {\bibfnamefont {R.~J.}\ \bibnamefont
  {Williams}},\ }\bibfield  {title} {\bibinfo {title} {Learning representations
  by back-propagating errors},\ }\href
  {https://api.semanticscholar.org/CorpusID:205001834} {\bibfield  {journal}
  {\bibinfo  {journal} {Nature}\ }\textbf {\bibinfo {volume} {323}},\ \bibinfo
  {pages} {533} (\bibinfo {year} {1986})}\BibitemShut {NoStop}%
\bibitem [{\citenamefont {Graves}\ \emph {et~al.}(2013)\citenamefont {Graves},
  \citenamefont {Mohamed},\ and\ \citenamefont {Hinton}}]{Alex2013DeepRNN}%
  \BibitemOpen
  \bibfield  {author} {\bibinfo {author} {\bibfnamefont {A.}~\bibnamefont
  {Graves}}, \bibinfo {author} {\bibfnamefont {A.-r.}\ \bibnamefont
  {Mohamed}},\ and\ \bibinfo {author} {\bibfnamefont {G.}~\bibnamefont
  {Hinton}},\ }\bibfield  {title} {\bibinfo {title} {Speech recognition with
  deep recurrent neural networks},\ }in\ \href
  {https://doi.org/10.1109/ICASSP.2013.6638947} {\emph {\bibinfo {booktitle}
  {2013 IEEE International Conference on Acoustics, Speech and Signal
  Processing}}}\ (\bibinfo {year} {2013})\ pp.\ \bibinfo {pages}
  {6645--6649}\BibitemShut {NoStop}%
\bibitem [{\citenamefont {Lecun}\ \emph {et~al.}(1998)\citenamefont {Lecun},
  \citenamefont {Bottou}, \citenamefont {Bengio},\ and\ \citenamefont
  {Haffner}}]{1Lecun1998CNN_LeNet}%
  \BibitemOpen
  \bibfield  {author} {\bibinfo {author} {\bibfnamefont {Y.}~\bibnamefont
  {Lecun}}, \bibinfo {author} {\bibfnamefont {L.}~\bibnamefont {Bottou}},
  \bibinfo {author} {\bibfnamefont {Y.}~\bibnamefont {Bengio}},\ and\ \bibinfo
  {author} {\bibfnamefont {P.}~\bibnamefont {Haffner}},\ }\bibfield  {title}
  {\bibinfo {title} {Gradient-based learning applied to document recognition},\
  }\href {https://doi.org/10.1109/5.726791} {\bibfield  {journal} {\bibinfo
  {journal} {Proceedings of the IEEE}\ }\textbf {\bibinfo {volume} {86}},\
  \bibinfo {pages} {2278} (\bibinfo {year} {1998})}\BibitemShut {NoStop}%
\bibitem [{\citenamefont {Dosovitskiy}\ \emph {et~al.}(2021)\citenamefont
  {Dosovitskiy}, \citenamefont {Beyer}, \citenamefont {Kolesnikov},
  \citenamefont {Weissenborn}, \citenamefont {Zhai}, \citenamefont
  {Unterthiner}, \citenamefont {Dehghani}, \citenamefont {Minderer},
  \citenamefont {Heigold}, \citenamefont {Gelly}, \citenamefont {Uszkoreit},\
  and\ \citenamefont {Houlsby}}]{Dosovitskiy2021ViT}%
  \BibitemOpen
  \bibfield  {author} {\bibinfo {author} {\bibfnamefont {A.}~\bibnamefont
  {Dosovitskiy}}, \bibinfo {author} {\bibfnamefont {L.}~\bibnamefont {Beyer}},
  \bibinfo {author} {\bibfnamefont {A.}~\bibnamefont {Kolesnikov}}, \bibinfo
  {author} {\bibfnamefont {D.}~\bibnamefont {Weissenborn}}, \bibinfo {author}
  {\bibfnamefont {X.}~\bibnamefont {Zhai}}, \bibinfo {author} {\bibfnamefont
  {T.}~\bibnamefont {Unterthiner}}, \bibinfo {author} {\bibfnamefont
  {M.}~\bibnamefont {Dehghani}}, \bibinfo {author} {\bibfnamefont
  {M.}~\bibnamefont {Minderer}}, \bibinfo {author} {\bibfnamefont
  {G.}~\bibnamefont {Heigold}}, \bibinfo {author} {\bibfnamefont
  {S.}~\bibnamefont {Gelly}}, \bibinfo {author} {\bibfnamefont
  {J.}~\bibnamefont {Uszkoreit}},\ and\ \bibinfo {author} {\bibfnamefont
  {N.}~\bibnamefont {Houlsby}},\ }\bibfield  {title} {\bibinfo {title} {An
  image is worth 16x16 words: Transformers for image recognition at scale},\
  }\href {https://arxiv.org/abs/2010.11929} {\bibfield  {journal} {\bibinfo
  {journal} {arXiv:2010.11929}\ } (\bibinfo {year} {2021})}\BibitemShut
  {NoStop}%
\bibitem [{\citenamefont {He}\ \emph {et~al.}(2015{\natexlab{a}})\citenamefont
  {He}, \citenamefont {Zhang}, \citenamefont {Ren},\ and\ \citenamefont
  {Sun}}]{he2015ResnetV1}%
  \BibitemOpen
  \bibfield  {author} {\bibinfo {author} {\bibfnamefont {K.}~\bibnamefont
  {He}}, \bibinfo {author} {\bibfnamefont {X.}~\bibnamefont {Zhang}}, \bibinfo
  {author} {\bibfnamefont {S.}~\bibnamefont {Ren}},\ and\ \bibinfo {author}
  {\bibfnamefont {J.}~\bibnamefont {Sun}},\ }\bibfield  {title} {\bibinfo
  {title} {Deep residual learning for image recognition},\ }\href
  {https://doi.org/10.48550/arXiv.1512.03385} {\bibfield  {journal} {\bibinfo
  {journal} {arXiv:1512.03385}\ } (\bibinfo {year}
  {2015}{\natexlab{a}})}\BibitemShut {NoStop}%
\bibitem [{\citenamefont {He}\ \emph {et~al.}(2016)\citenamefont {He},
  \citenamefont {Zhang}, \citenamefont {Ren},\ and\ \citenamefont
  {Sun}}]{he2016ResnetV2}%
  \BibitemOpen
  \bibfield  {author} {\bibinfo {author} {\bibfnamefont {K.}~\bibnamefont
  {He}}, \bibinfo {author} {\bibfnamefont {X.}~\bibnamefont {Zhang}}, \bibinfo
  {author} {\bibfnamefont {S.}~\bibnamefont {Ren}},\ and\ \bibinfo {author}
  {\bibfnamefont {J.}~\bibnamefont {Sun}},\ }\bibfield  {title} {\bibinfo
  {title} {Identity mappings in deep residual networks},\ }\href
  {https://doi.org/10.48550/arXiv.1603.05027} {\bibfield  {journal} {\bibinfo
  {journal} {arXiv:1603.05027}\ } (\bibinfo {year} {2016})}\BibitemShut
  {NoStop}%
\bibitem [{\citenamefont {Zhang}\ \emph {et~al.}(2023)\citenamefont {Zhang},
  \citenamefont {Wan},\ and\ \citenamefont {Yao}}]{zhang2023ADMC}%
  \BibitemOpen
  \bibfield  {author} {\bibinfo {author} {\bibfnamefont {S.-X.}\ \bibnamefont
  {Zhang}}, \bibinfo {author} {\bibfnamefont {Z.-Q.}\ \bibnamefont {Wan}},\
  and\ \bibinfo {author} {\bibfnamefont {H.}~\bibnamefont {Yao}},\ }\bibfield
  {title} {\bibinfo {title} {Automatic differentiable monte carlo: Theory and
  application},\ }\href {https://doi.org/10.1103/PhysRevResearch.5.033041}
  {\bibfield  {journal} {\bibinfo  {journal} {Phys. Rev. Res.}\ }\textbf
  {\bibinfo {volume} {5}},\ \bibinfo {pages} {033041} (\bibinfo {year}
  {2023})}\BibitemShut {NoStop}%
\bibitem [{\citenamefont {Sinibaldi}\ \emph {et~al.}(2023)\citenamefont
  {Sinibaldi}, \citenamefont {Giuliani}, \citenamefont {Carleo},\ and\
  \citenamefont {Vicentini}}]{Sinibaldi2023unbiasingtime}%
  \BibitemOpen
  \bibfield  {author} {\bibinfo {author} {\bibfnamefont {A.}~\bibnamefont
  {Sinibaldi}}, \bibinfo {author} {\bibfnamefont {C.}~\bibnamefont {Giuliani}},
  \bibinfo {author} {\bibfnamefont {G.}~\bibnamefont {Carleo}},\ and\ \bibinfo
  {author} {\bibfnamefont {F.}~\bibnamefont {Vicentini}},\ }\bibfield  {title}
  {\bibinfo {title} {Unbiasing time-dependent {V}ariational {M}onte {C}arlo by
  projected quantum evolution},\ }\href
  {https://doi.org/10.22331/q-2023-10-10-1131} {\bibfield  {journal} {\bibinfo
  {journal} {{Quantum}}\ }\textbf {\bibinfo {volume} {7}},\ \bibinfo {pages}
  {1131} (\bibinfo {year} {2023})}\BibitemShut {NoStop}%
\bibitem [{\citenamefont {Gravina}\ \emph {et~al.}(2025)\citenamefont
  {Gravina}, \citenamefont {Savona},\ and\ \citenamefont
  {Vicentini}}]{gravina2025neuralprojectedquantumdynamics}%
  \BibitemOpen
  \bibfield  {author} {\bibinfo {author} {\bibfnamefont {L.}~\bibnamefont
  {Gravina}}, \bibinfo {author} {\bibfnamefont {V.}~\bibnamefont {Savona}},\
  and\ \bibinfo {author} {\bibfnamefont {F.}~\bibnamefont {Vicentini}},\
  }\bibfield  {title} {\bibinfo {title} {Neural projected quantum dynamics: a
  systematic study},\ }\href {https://arxiv.org/abs/2410.10720} {\bibfield
  {journal} {\bibinfo  {journal} {arXiv:2410.10720}\ } (\bibinfo {year}
  {2025})}\BibitemShut {NoStop}%
\bibitem [{\citenamefont {Jónsson}\ \emph {et~al.}(2018)\citenamefont
  {Jónsson}, \citenamefont {Bauer},\ and\ \citenamefont
  {Carleo}}]{jonsson2018neuralnetworkstatesclassicalsimulation}%
  \BibitemOpen
  \bibfield  {author} {\bibinfo {author} {\bibfnamefont {B.}~\bibnamefont
  {Jónsson}}, \bibinfo {author} {\bibfnamefont {B.}~\bibnamefont {Bauer}},\
  and\ \bibinfo {author} {\bibfnamefont {G.}~\bibnamefont {Carleo}},\
  }\bibfield  {title} {\bibinfo {title} {Neural-network states for the
  classical simulation of quantum computing},\ }\href
  {https://arxiv.org/abs/1808.05232} {\bibfield  {journal} {\bibinfo  {journal}
  {arXiv:1808.05232}\ } (\bibinfo {year} {2018})}\BibitemShut {NoStop}%
\bibitem [{\citenamefont {Choo}\ \emph {et~al.}(2019)\citenamefont {Choo},
  \citenamefont {Neupert},\ and\ \citenamefont
  {Carleo}}]{Choo2019_CNN_ComplexValued}%
  \BibitemOpen
  \bibfield  {author} {\bibinfo {author} {\bibfnamefont {K.}~\bibnamefont
  {Choo}}, \bibinfo {author} {\bibfnamefont {T.}~\bibnamefont {Neupert}},\ and\
  \bibinfo {author} {\bibfnamefont {G.}~\bibnamefont {Carleo}},\ }\bibfield
  {title} {\bibinfo {title} {Two-dimensional frustrated
  ${J}_{1}\text{\ensuremath{-}}{J}_{2}$ model studied with neural network
  quantum states},\ }\href {https://doi.org/10.1103/PhysRevB.100.125124}
  {\bibfield  {journal} {\bibinfo  {journal} {Phys. Rev. B}\ }\textbf {\bibinfo
  {volume} {100}},\ \bibinfo {pages} {125124} (\bibinfo {year}
  {2019})}\BibitemShut {NoStop}%
\bibitem [{\citenamefont {Kingma}\ and\ \citenamefont
  {Ba}(2017)}]{Kingma2015Adam}%
  \BibitemOpen
  \bibfield  {author} {\bibinfo {author} {\bibfnamefont {D.~P.}\ \bibnamefont
  {Kingma}}\ and\ \bibinfo {author} {\bibfnamefont {J.}~\bibnamefont {Ba}},\
  }\bibfield  {title} {\bibinfo {title} {Adam: A method for stochastic
  optimization},\ }\href {https://doi.org/10.48550/arXiv.1412.6980} {\bibfield
  {journal} {\bibinfo  {journal} {arXiv:1412.6980}\ } (\bibinfo {year}
  {2017})}\BibitemShut {NoStop}%
\bibitem [{\citenamefont {Liu}\ \emph {et~al.}(2022)\citenamefont {Liu},
  \citenamefont {Gong}, \citenamefont {Li}, \citenamefont {Poilblanc},
  \citenamefont {Chen},\ and\ \citenamefont
  {Gu}}]{Zheng-ChengGu2022_TensorNetwork+PEPS}%
  \BibitemOpen
  \bibfield  {author} {\bibinfo {author} {\bibfnamefont {W.-Y.}\ \bibnamefont
  {Liu}}, \bibinfo {author} {\bibfnamefont {S.-S.}\ \bibnamefont {Gong}},
  \bibinfo {author} {\bibfnamefont {Y.-B.}\ \bibnamefont {Li}}, \bibinfo
  {author} {\bibfnamefont {D.}~\bibnamefont {Poilblanc}}, \bibinfo {author}
  {\bibfnamefont {W.-Q.}\ \bibnamefont {Chen}},\ and\ \bibinfo {author}
  {\bibfnamefont {Z.-C.}\ \bibnamefont {Gu}},\ }\bibfield  {title} {\bibinfo
  {title} {Gapless quantum spin liquid and global phase diagram of the
  spin-$1/2$ $j_1$-$j_2$ square antiferromagnetic {Heisenberg} model},\ }\href
  {https://doi.org/https://doi.org/10.1016/j.scib.2022.03.010} {\bibfield
  {journal} {\bibinfo  {journal} {Science Bulletin}\ }\textbf {\bibinfo
  {volume} {67}},\ \bibinfo {pages} {1034} (\bibinfo {year}
  {2022})}\BibitemShut {NoStop}%
\bibitem [{\citenamefont {Marshall}(1955)}]{1955MarshallSign}%
  \BibitemOpen
  \bibfield  {author} {\bibinfo {author} {\bibfnamefont {W.}~\bibnamefont
  {Marshall}},\ }\bibfield  {title} {\bibinfo {title} {Antiferromagnetism},\
  }\href@noop {} {\bibfield  {journal} {\bibinfo  {journal} {Proc. R. Soc.
  Lond. A}\ }\textbf {\bibinfo {volume} {232}},\ \bibinfo {pages} {48}
  (\bibinfo {year} {1955})}\BibitemShut {NoStop}%
\bibitem [{\citenamefont {He}\ \emph {et~al.}(2015{\natexlab{b}})\citenamefont
  {He}, \citenamefont {Zhang}, \citenamefont {Ren},\ and\ \citenamefont
  {Sun}}]{he2015KaimingInitilization}%
  \BibitemOpen
  \bibfield  {author} {\bibinfo {author} {\bibfnamefont {K.}~\bibnamefont
  {He}}, \bibinfo {author} {\bibfnamefont {X.}~\bibnamefont {Zhang}}, \bibinfo
  {author} {\bibfnamefont {S.}~\bibnamefont {Ren}},\ and\ \bibinfo {author}
  {\bibfnamefont {J.}~\bibnamefont {Sun}},\ }\bibfield  {title} {\bibinfo
  {title} {Delving deep into rectifiers: Surpassing human-level performance on
  imagenet classification},\ }\href {https://doi.org/10.48550/arXiv.1502.01852}
  {\bibfield  {journal} {\bibinfo  {journal} {arXiv:1502.01852}\ } (\bibinfo
  {year} {2015}{\natexlab{b}})}\BibitemShut {NoStop}%
\bibitem [{\citenamefont {Sinibaldi}\ \emph {et~al.}(2025)\citenamefont
  {Sinibaldi}, \citenamefont {Hendry}, \citenamefont {Vicentini},\ and\
  \citenamefont {Carleo}}]{sinibaldi2025timedependentneuralgalerkinmethod}%
  \BibitemOpen
  \bibfield  {author} {\bibinfo {author} {\bibfnamefont {A.}~\bibnamefont
  {Sinibaldi}}, \bibinfo {author} {\bibfnamefont {D.}~\bibnamefont {Hendry}},
  \bibinfo {author} {\bibfnamefont {F.}~\bibnamefont {Vicentini}},\ and\
  \bibinfo {author} {\bibfnamefont {G.}~\bibnamefont {Carleo}},\ }\bibfield
  {title} {\bibinfo {title} {Time-dependent neural {G}alerkin method for
  quantum dynamics},\ }\href {https://arxiv.org/abs/2412.11778} {\bibfield
  {journal} {\bibinfo  {journal} {arXiv:2412.11778}\ } (\bibinfo {year}
  {2025})}\BibitemShut {NoStop}%
\bibitem [{Note1()}]{Note1}%
  \BibitemOpen
  \bibinfo {note} {The code supporting the findings of this study is openly
  available in a GitHub repository at https://github.com/QTMEC-RUC/nqs-lanczos
  (will be publicly accessible upon publication of the paper).}\BibitemShut
  {Stop}%
\end{thebibliography}%

\end{document}